\documentclass[final,1p,times]{elsarticle}

\usepackage{amssymb}
\usepackage{array}
\newcolumntype{P}[1]{>{\centering\arraybackslash}p{#1}}
\usepackage{lineno}
\usepackage[T1]{fontenc}
\usepackage{amsthm}
\usepackage{amsmath}

\newtheorem{mydef}{Definition}
\newcommand*{\bigot}{\raisebox{0.20ex}{\scalebox{0.6}{$\bigotimes$}}}
\usepackage{subcaption}
\usepackage[colorlinks=true,linkcolor=blue]{hyperref}%

\begin{document}

\begin{frontmatter}



\title{Knowledge Graph Completion: A Bird's Eye View on Knowledge Graph Embeddings, Software Libraries, Applications and Challenges}
\author[add1]{Satvik Garg}
\address[add1]{Department of Computer Science, Jaypee University of Information Technology, Solan, India}
\ead{satvikgarg27@gmail.com}
\author[add2]{Dwaipayan Roy\corref{cor1}}
\address[add2]{Department of Computational and Data Sciences, Indian Institute of Science, Education and Research, Kolkata, India}
\ead{dwaipayan.roy@iiserkol.ac.in}
\cortext[cor1]{Corresponding author}



\begin{abstract}
In recent years, Knowledge Graph (KG) development has attracted significant researches considering the applications in web search, relation prediction, natural language processing, information retrieval, question answering to name a few. However, often KGs are incomplete due to which Knowledge Graph Completion (KGC) has emerged as a sub-domain of research to automatically track down the missing connections in a KG.
Numerous strategies have been suggested to work out the KGC dependent on different representation procedures intended to embed triples into a low-dimensional vector space. Given the difficulties related to KGC, researchers around the world are attempting to comprehend the attributes of the problem statement. This study intends to provide an overview of knowledge bases combined with different challenges and their impacts. We discuss existing KGC approaches, including the state-of-the-art Knowledge Graph Embeddings (KGE), not only on static graphs but also for the latest trends such as multimodal, temporal, and uncertain knowledge graphs. In addition, reinforcement learning techniques are reviewed to model complex queries as a link prediction problem. Subsequently, we explored popular software packages for model training and examine open research challenges that can guide future research.
\end{abstract}


\end{frontmatter}


\section{Introduction}
The concept of  Knowledge Graphs (KG) was proposed by Google in 2012 to utilize semantic information in web search to enhance the performance of web crawlers and upgrade the experience of clients \cite{hogan2021knowledge}. The Knowledge Graphs are based on numerous information retrieval frameworks that obtain admittance to organized information and are utilized to distinguish and disambiguate elements in text, advance query response with semantically organized outlines, and give links to related entities in experimental search. Leveraging real world information in data frameworks is one of the significant advancements in automation \cite{mycin1976computer}. Representation of data and logic inspired by human critical thinking expected to present data or information to secure, improve the ability to deal with complex questions and have drawn in incredible scholarly thought and professions \cite{hogan2021knowledge} \cite{nickel2015review} \cite{ji2021survey} \cite{chen2020knowledge}. 

Knowledge graph is a graph-based data representation modality consisting of binary relationships and labeled edges. It comprises real-world triplets, where each triplet or fact $(e1, r, e2)$ addresses a connection $r$ between head entity $e1$ and tail entity $e2$. They are often called as multi relational graphs where each node and edge represents an element (or entity) and a relation, respectively. The relations helps to connect nodes to encode different links separately. The entities can be addressed as things of real world knowledge such as film, person, city, country, place to name a few. An example is shown in figure 1, a relation \textit{`Friends with'} connects person entity type, and the connection type \textit{`works in'} represents the relationship between entity type of person and organization.

The knowledge graphs are important for many applicative use cases like social networks, web-based collaborative knowledge bases like DBpedia \cite{bizer2009dbpedia}, and in healthcare when trying to model protein-protein interaction networks or genetic information \cite{mohamed2020discovering}. They are also useful for Natural Language Processing (NLP) applications like entity recognition \cite{luo2015joint}, entity linking \cite{shen2014entity}, dialogue systems \cite{he2017learning}, semantic parsing \cite{krishnamurthy2012weakly}, information retrieval \cite{reinanda2020knowledge} and question answering systems \cite{fader2014open}. Most KG's are available online and open sourced ranging from domain specific KG's such as GeneOntology \cite{gene2017expansion} and for general purposes such as YAGO \cite{suchanek2008yago}, FreeBase \cite{bollacker2007freebase}, DBpedia \cite{bizer2009dbpedia}, WordNet \cite{miller1998wordnet}, NELL \cite{carlson2010toward}. Knowledge Graphs are the result of automatic generation, in some cases from mining web pages like GDELT \cite{ward2013comparing} and craft source operations like WIKIDATA \cite{vrandevcic2014wikidata}. Commercial KG's are pretty common in applications like search engines. Examples of commercial KG include Facebook Open Graph, Microsoft Satori, Yahoo Spark and Google Knowledge Graph \cite{overland2011facebook}.

\begin{figure}[htbp]
\centerline{\includegraphics[width=8cm, height=5.5cm]{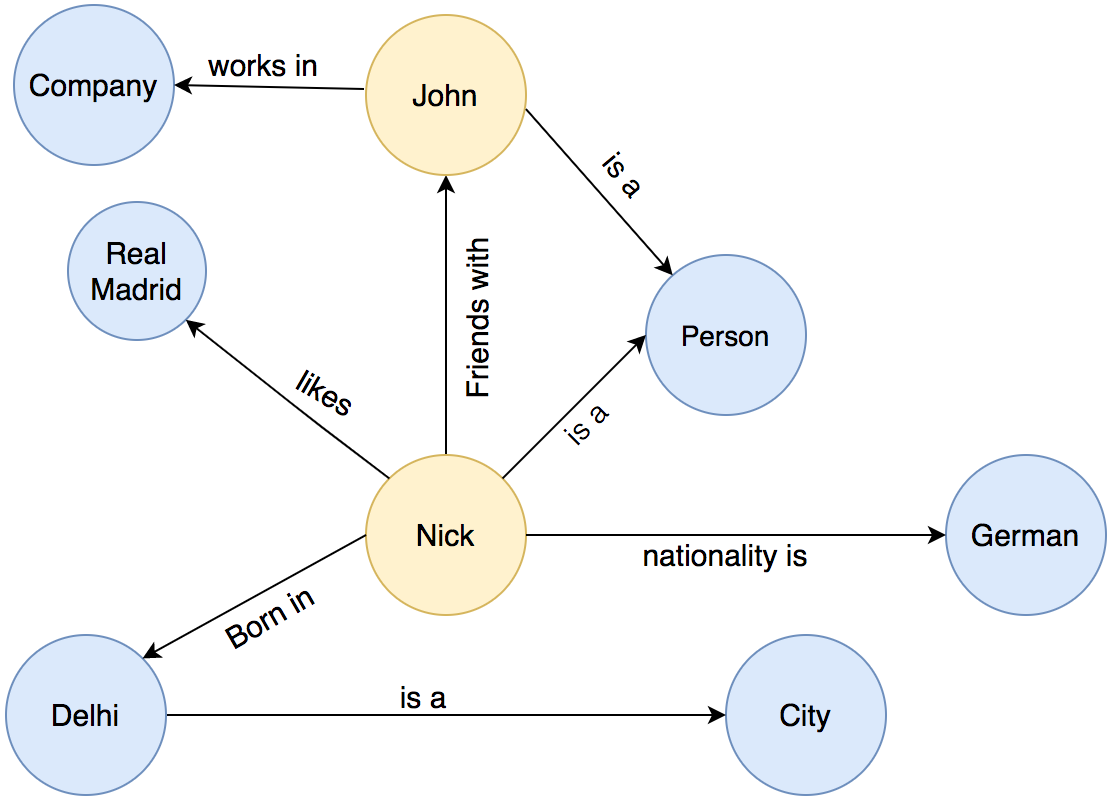}}
\caption{An example of hypothetical knowledge graph.}
\end{figure}

Knowledge graphs are often generated automatically, have missing edges, and may not be completely comprehensive. An emerging issue arise when huge KG's, for example, DBpedia and Freebase contain many facts or triples on real world knowledge and are long way from complete \cite{west2014knowledge} \cite{krompass2015type}. In Freebase, it has been shown that about 70\% of people have missing birth status, while 75\% miss identification and 95\% have no information on their parents \cite{west2014knowledge}. In DBpedia, like Freebase, around 65\% of people do not have any birthplace knowledge, and 60\% of scientists miss information on their study area \cite{krompass2015type}. Some reasons are introduced as to why KG's have flaws of its own and are mainly fragmentary. First, recognizing billions of reals on human knowledge is not scalable. In addition, accurate information and data are dynamically advancing, making it challenging to build complete and suitable KGs. A link prediction problem is presented as an examination area named Knowledge Graph Completion (KGC) \cite{chen2020knowledge} to combat all the previously mentioned issues and helps to anticipate the missing connection and knowledge between entities or facts. An example is shown in figure 2 to comprehend the topic of link prediction easier to understand. Strong lines address total or existing relationships, while red dotted lines represent potential relationships that may extend the knowledge graph completeness.

Leveraging automation techniques in graphs can be helpful as graphs are enormous and carries a lot of knowledge. Problems such as link prediction and triple classification are used for graph completion, content recommendation, and question-answer systems. Triple classification determines whether a link missing is true or false \cite{shijia2016knowledge} and is a binary classification task. There are other valuable areas of study like collective node classification \cite{bilgic2007combining} and link-based clustering \cite{saeedi2018using} used to assign a label to two nodes according to its topology and structure of graph, which is useful for customer segmentation []. The mapping of copied items can easily be dissected utilizing the concept of Entity Matching (EM) \cite{pershina2015holistic}, which is principal to link information of similar real world entities and helps in knowledge refining []. It is an active research area and considered a significant advance to perform downstream tasks like entity linking, triple classification, and many more. 

This work focused merely on knowledge graph completion (KGC) to generate the ranking of missing relationships. It involves two subtasks: entity ranking and relationship prediction. The entity ranking problem is responsible for discovering missing elements, given $?$ as a missing connection or entity, anticipate $e1$ given $(?, r, e2)$ or $e2$ given $(e1, r, ?)$ in contrast to relation prediction problem to rank the missing connection, foreseeing $r$ given $(e1, ?, e2)$.  The main objective consists in using the facts and relations given in KG as an aid to increase the probability of finding the missing elements. For example, the relationship between a person element and a country element can be easily dissected by knowing a person's neighborhood and the country that city is located in. 

\begin{figure}
\centerline{\includegraphics[width=8cm, height=5.5cm]{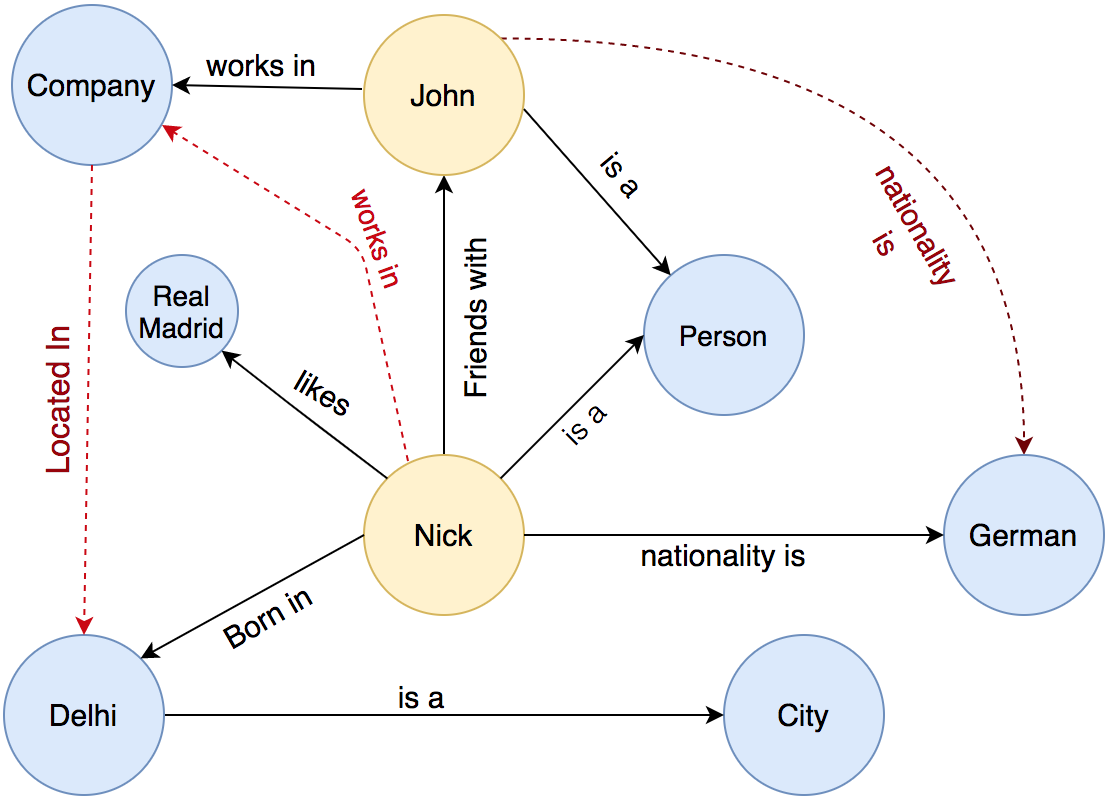}}
\caption{An example of hypothetical knowledge graph for knowledge graph completion. Red dotted line represents potential links.}
\end{figure}

Several approaches exist to tackle the challenging problem of link prediction []. The decomposition based method maps the given entities, links them to tensors, and provides expected semantic data \cite{chang2014typed}. The path based approaches, including earliest random walks and path ranking algorithm (PRA) \cite{lao2011random}, potentially include a path between destinations through a sequence of edges. In particular, these models experience low proficiency, versatility and may consist of numerous parameters making models computationally costly to prepare []. Knowledge Graph Embeddings (KGE) have been proposed to handle the aforementioned challenges and acquired huge consideration recently \cite{wang2017knowledge}. The aim is to extract meaningful knowledge, i.e., entities and relations from a given knowledge graph, and install them into a continuous low dimensional vector space to perform downstream tasks like KGC, triple classification, entity resolution, collective node classification, and so forth. More importantly, it works on the intricacy and improves scalability while safeguarding the intrinsic construction of graphs. 

The KGE models are mostly categorized into three different types of models namely, translation, semantic, and neural network based. Models such as TransE \cite{bordes2013translating}, TransH \cite{wang2014knowledge}, TransR \cite{lin2015learning}, TransD \cite{ji2015knowledge}, TranSparse \cite{ji2016knowledge} are common examples of translation based approaches. DistMult \cite{yang2014embedding}, RESCAL \cite{nickel2011three}, CompleX \cite{trouillon2016complex} belong to the semantic-based methods. These models are good yet fail to provide deeper semantics, ignoring hierarchical associations. Contrary to previous techniques, neural based methods such as ConvKB \cite{nguyen2017novel}, ConvE \cite{dettmers2018convolutional}, HypER \cite{balavzevic2019hypernetwork} generate state of the art (SOTA) performance by employing deep learning structures assessing temporal features, path, and structural information that helps produce better embeddings for downstream tasks.

\subsection{Related Works and Contributions}
There exists several surveys available for Knowledge Graph Completion. Wang et al. \cite{wang2021survey} provide a review of KGE techniques for link prediction and compared the analysis between the performance of different models. Similar work was done by Dai et al. \cite{dai2020survey} to conduct a thorough analysis by manually training the KGE models. They likewise surveyed the works that leverage extra semantic data dependent on text-based features and relation paths. Both surveys focused more on overseeing a comparative analysis by conducting experiments on standard datasets. When looking at research papers on KGE for link prediction, a mere disconnection between the state-of-the-art techniques reported in articles and the techniques actually employed in some related applications has been noticed []. So, we focused more on delivering practical post-processing techniques for comparative analysis such as hyper parameter tuning, calibration techniques \cite{safavi2020evaluating} and Neighborhood Inconsistency Matrix \cite{wangneighborhood} that helps connect fellow researchers to identify the recent works with more dimensions. Taking one step forward, a brief overview of graph representation learning methods based on traditional statistical learning methods, graph-based features methods, graph neural networks \cite{scarselli2008graph} are also discussed in this work.

Papers by Chen et al. \cite{chen2020knowledge} and Ji et al. \cite{ji2021survey} are most up-to-date survey papers on Knowledge Graph Completion and Representation. Chen et al. \cite{chen2020knowledge} briefly discussed the advantages and disadvantages of KGE models and provides an overview of KGC based on Network Representation Learning. Ji et al. \cite{ji2021survey} reviewed the knowledge graphs for representation, acquisition, and their applications. They covered the broad spectrum of knowledge graphs in terms of KGC, relation extraction, and entity classification. However, both these papers lack the studies to leverage multimodal Knowledge Graphs. We, therefore, attempt to give a survey of methods on KGE strategies, including not only on the single facts alone but also on KGs that further leverage multimodal information such as images, text, and timestamps \cite{pezeshkpour2018embedding}. More importantly, we present how learned embeddings can be applied to an advantage over a wide assortment of applications. Rossi et al. \cite{rossi2021knowledge} ordered models into three classifications: mathematical models, tensor decomposition models, and neural network models. They chose standard models for point-by-point depiction, experimental result comparison, and analysis for these three classes. Nonetheless, there is no general grouping and outline of the KGE models proposed lately in this paper, and the chosen models are not many, which cannot cover a wide range of KGE models. In addition, reinforcement learning techniques to model complex QA systems are reviewed in Section 4.

Real-world knowledge comprises multimodality such as images, text and timestamps. Leveraging literals in KG is a challenging task \cite{pezeshkpour2018embedding}. Gesese et al. \cite{gesese2019survey} conducted a survey on KGE for multimodal knowledge graphs, in which they covered models such as KBLRN \cite{garcia2017kblrn}, LiteralE \cite{kristiadi2019incorporating}, and many more. An overview of existing techniques related to multimodality, including temporal and uncertain knowledge graphs, is also conducted in our survey providing readers a comprehensive view on KGC. Most of the works focus on providing applications related to Knowledge Graphs. Abu-Salih et al. \cite{abu2021domain} surveyed domain-specific KG to give readers a thorough review of the state-of-the-art approaches drawn from academic works relevant to different knowledge domains. Zou et al. \cite{zou2020survey} conducted a bird's eye view on applications stemming from different domains like Question answering and recommendation systems. Taking their work forward, we also included the applications of knowledge graphs in the context of COVID-19, healthcare informatics, drug discovery, human resource and knowledge protection.

To the best of our knowledge, no past examination gives an orderly survey of the software libraries that revolves around Knowledge graph embeddings. 
As a result, unlike other related works that focus on KG development, this work aims to provide the first survey on these software ecosystems available for KGE training to perform downstream tasks. 
This paper likewise points to significant progress in applying knowledge graphs to available open research challenges, such as robustness, interpretability, scalability, and few-shot learning, and many more.

We provide readers with a bird's eye view on understanding the concepts required for the KGE model, and include a survey of SOTA KGE methods with the most recent patterns. This survey additionally goes deeper into the flow of KGE and provides a full-scale view of the KGE pipeline, including loss functions, scoring functions, negative generation, evaluation metrics, and auxiliary information for downstream tasks such as Link prediction. 

The following are the notable contributions:
\vspace{-0.5em}
\begin{itemize}
\setlength\itemsep{-0.3em}
  \item Apparently, this is one of few endeavors to provide a beginner-friendly comprehensive review that covers all aspects related to knowledge graph completion. The scoring models are separated into three general classifications: translation-based, semantic matching based, and neural-based to make this review readers friendly.
  \item A high perspective is given on traditional relational learning and graph representation learning techniques that help new researchers to effectively understand research work from both traditional and non-traditional perspectives.
  \item Each of the steps identified with the construction of the KGE model are explored to perform downstream tasks, including the scoring function, loss function, negative generation, and optimization.
  \item We examined the recent developments made in reinforcement learning techniques for KGC to infer complex queries.
  \item  The KGE based on utilizing real world knowledge including numeric  values, text, images, uncertain and temporal information, for link prediction is reviewed.
  \item A Comparative analysis of open software libraries is analyzed for training knowledge graphs to perform downstream tasks.
  \item We explored the open research challenges such as robustness, scalability, few shot learning, knowledge transfer, multi-path predictions that help direct future examination.
  \item Different use cases identified with KG are mentioned in the context of question answering, recommendation, information retrieval, COVID-19, etc., to help readers understand the real-world applications.
\end{itemize}

This study has nine segments and is organized as follows: Section 2 outlines traditional relational learning and graph representation learning techniques.  Section 3 examines existing knowledge graph embedding models for link prediction. Section 4 contains a comparative analysis of various methodologies for KGC. Section 5 review the reinforcement learning techniques for KGC. Section 6 incorporates techniques to leverage multimodality in knowledge graphs for KGC. Section 7 aims to give concise information on open-source software libraries for executing KGE models for link prediction. Section 7 examines a few applications identified with the Knowledge Graph. Section 8 contains open research challenges. Lastly, we conclude this study in Section 9.

\section{On Relational Learning and Representational Learning}

This section discusses an overview of traditional statistical relational learning and graph representation learning methods for the task of link prediction. This will help fellow researchers to develop a good understanding of problem statement from traditional and non-traditional perspectives.

\subsection{Traditional Relational Learning}

The realm of statistical relational learning is quite an established field \cite{popescul2003statistical}. There are several techniques proposed over the years that are widely used to predict new links from facts. However, the design logic and techniques used are quite different compared to the current state of art models. It is important to know that these methods exist, for example, similarity-based methods \cite{yu2017similarity}, inductive logical programming \cite{lavrac1994inductive}, rule mining \cite{galarraga2013mining}, graphical models such as bayesian and markov logic networks \cite{sucar2015probabilistic}.

\subsubsection{Similarity Based Methods}

Techniques that incorporate similarity based features originally centered around the topological construction of graphs are the most straightforward and traditional link prediction practice \cite{yu2017similarity}. Similarity based methods are broadly utilized for predicting the missing links in graphs that consist of only one connection, for example; in science (associations between protein), social networks (friend recommendation), web mining (hyperlinks between web destinations). It works by assigning a comparability score between node pairs using the underlying topology of the graph. The tendency behind this approach is that entities are probably being connected if they are similar and can be measured by the locality of nodes or by the presence of walk between nodes. It can be researched under three principal classifications: local \cite{kaushik2002exploiting}, global \cite{anand2014exploring}, and quasi local methodologies \cite{liu2018network}. Local similarity based techniques, for example, Academic Adar index \cite{adamic2003friends}, Jaccard Index \cite{jaccard1901etude}, Salton Index \cite{chowdhury2010introduction}, Common Neighbors \cite{newman2001clustering} infer the closeness or similarity of entities from their outright number of neighbors. They are quick to implement for single connections and scale well to enormous information graphs as their calculation relies just upon the neighborhood of the participating entities. They can however be too restricted to even consider catching significant patterns and may not show long range dependencies. As opposed to local, the global similarity methods \cite{anand2014exploring} utilize the entire structure of the graph to rank the similitude between nodes even if the nodes are slightly further. In spite of the fact that the entire topology of the graph gives greater adaptability in prediction, it likewise expands the computational time since a fusion of all paths between nodes is incorporated. Some common examples include Katz Index \cite{katz1953new}, SimRank \cite{jeh2002simrank}, Rooted PageRank \cite{wang2015link}, Random Walk with Restart \cite{lu2010link}. The compromise between the proficiency of the data with respect to the entire graph topological design (global methodologies) and reduced time based methods (local similarity) approaches have brought about the rise of quasi local similarity based methods \cite{liu2018network}. Semi neighborhood similarity examples include Local Path Index \cite{zhou2009predicting}, FriendLink Index \cite{papadimitriou2012fast}, Superposed and Local Random Walks \cite{liu2010link}. It attempts to adjust computational logic and precision by determining the likeness of entities from random walks and paths of limited length. A portion of these methods may leverage the entire topology of the graph still the complexity is lesser than global similarity based methods.  

\subsubsection{Rule Induction and Reasoning}

Taking in rules from KGs is a significant errand for link prediction, cleaning and classification. \textit{"Rules over graphs are of the structure head $\leftarrow$ body, where head is a binary atom and body is a combination of possibly negated binary and unary atoms"} \cite{stepanova2018rule}. It can be utilized to recognize noticeable examples from KGs and cast them as Horn rules. The objective of Inductive Logic Programming (ILP) \cite{lavrac1994inductive} is to sum up individual examples within the sight of background information by building speculations about unseen occurrences. Background information is considered as a cluster of triples or facts over different relations and rules that can be utilized to actuate the meaning of a logic program. The most general task in ILP is the task of learning sensible meanings of relationships. In particular, the conventional ILP assignment of gaining from both positive and negative instances is called learning from entailment. ALEPH is an ILP method that takes in rules using inverse entailment \cite{srinivasan2001aleph}. A portion of the common models for Horn and non monotonic principle enlistment are CIGOL \cite{muggleton1988machine}, CLINT \cite{de1991clint}, LINUS \cite{dvzeroski1991learning}, GOLEM \cite{cropper2019learning}, ILASP \cite{law2015ilasp}, ILED \cite{katzouris2015incremental}. Most of the traditional existing style standard rule induction strategies referenced above accept that the given information to which the rules are actuated is absolute, representative and precise. Accordingly, they depend on notions of closed world assumptions and are intended to mine rule hypotheses that fulfill inductive learning from instances. However, KG are exceptionally deficient, biased and error prone implying that the assignment of initiating an ideal rule set from a KG is ordinarily unworkable. Therefore, with respect to KG, one regularly aims to remove some generalities from the information, which is generally not true, when viewed as rules that a substantial portion of the confirmed facts are inferred.

The most conspicuous instances of such frameworks (rule mining) that are explicitly custom fitted towards prompting Horn rules from KGs are AMIE \cite{galarraga2015fast} and RDF2Rules \cite{wang2015rdf2rules}. AMIE receives the PCA proportion of certainty and assembles rules in a hierarchical design  beginning with rule heads like $\rightarrow ?x \hspace{1mm} nation \hspace{1mm} ?y$. For each rule on top of this structure (one for each edge name), three kinds of refinements are thought of adding dangling atom, instantiated atom and closing atom. It intends to augment another edge to the body of the rule. The execution of AMIE utilizes an assortment of procedures from the database region, which permit it to accomplish high scalability. While AMIE mines one rule at an instant, RDF2Rules parallelize this interaction by mining frequent predicate cycles (FPC). To separate FPCs, the RDF2Rules initially extract the frequent predicate paths (FPP). As soon as FPC are extracted, rules are then mined from them by picking a predicate to be in the rule head, and gathering the rest into its body. RDF2Rules is fit for representing unary predicates which are disregarded in AMIE for scalability issues. RDF2Rules plays out the standard extraction quicker than AMIE because of a viable pruning procedure utilized during mining FPC however the supported rule rationale is more prohibitive. An advantage of rule based mining approaches is that they are viably interpretable as some logical rules are given to the model.

\subsubsection{Probabilistic Graphical Models}

A probabilistic or graphical model \cite{sucar2015probabilistic} signifies the conditional independent structure for a graph between nodes. These models leverage the benefits of adaptable topological design, clear semantics and viable multi-data combination in managing complex issues. Graphical models infer a basic method to envision the construction of a probabilistic model and can be utilized to plan and propel new models. In a probabilistic graphical model, every node addresses a random variable, and the connections between them express probabilistic relations. The graph then, at that point, catches the paths by which the joint appropriation over the entirety of the nodes can be broken down into a product of elements relying just upon a subset of the random variables. The link prediction methods dependent on the graphical models generally utilizes Bayesian \cite{lu2017link} and Markov logic networks \cite{lao2010relational}.

Bayes' rule incorporates four different types of models namely, network evolution \cite{kashima2006parameterized}, stochastic models \cite{goldenberg2010survey}, structural models based on hierarchy \cite{ravasz2003hierarchical} and local probabilistic based methods \cite{wang2007local}. The main downside of a portion of these models is in effect sluggish and computationally expensive for enormous graphs \cite{al2011survey}. In contrast to a bayesian network, which incorporates Directed Acyclic Graph (DAG), the Relational Markov Network (RMN) is likewise proposed \cite{taskar2012discriminative} based on an undirected graph. RMN addresses two drawbacks of Bayesian networks that is they do not compel the graph to be non cyclic, which takes into account different conceivable representations of graphs. Moreover, they are more appropriate for discriminative modeling \cite{koller2007introduction}. There also exists other methods for solving the KGC. "The DAPER model is a DAG sort of probabilistic entity relationship model \cite{heckerman2007probabilistic}". The benefit of the DAPER model is to provide more expressivity than the previously mentioned approaches \cite{heckerman2004probabilistic}.

The manual feature extraction techniques that are explored in this study furnish a beginning stage to the efficient prediction of absent and future links accessible through learning the powerful characteristics in graphs. Among these powerful highlights for KGC, utilizing the structural features that can be mined from the graph is the stepping stone of all learning-based KGC techniques. The issue with manual feature extraction is generally that they are restricted in adaptability and for KG we need techniques that scale better given the size of the graph. Another drawback is that they have restricted model power unlike KG Embeddings and are not differentiable and cannot utilize current GPU designs with SGD learning. Other than the topological qualities, some machine learning based models may utilize the nodes with the domain explicit characteristics, alluded to as the proximity and accumulated features \cite{wang2016structural} \cite{tang2015line}.

\subsection{Graph Representation Learning}
While traditional machine learning models for KGC rely on hand-crafted feature engineering (as shown in figure 3), advances in graph representation learning (GRL) models have led to the introduction of automatically generated feature encoders, which prevent hand-designed features that forestall hand-designed attributes \cite{hamilton2020graph}. In simple words, GRL represents an area of study to apply machine learning on graphs, but avoid extracting features manually as they are difficult and time consuming on graphs. 

\begin{figure}[htbp]
\centerline{\includegraphics[width=8cm, height=3cm]{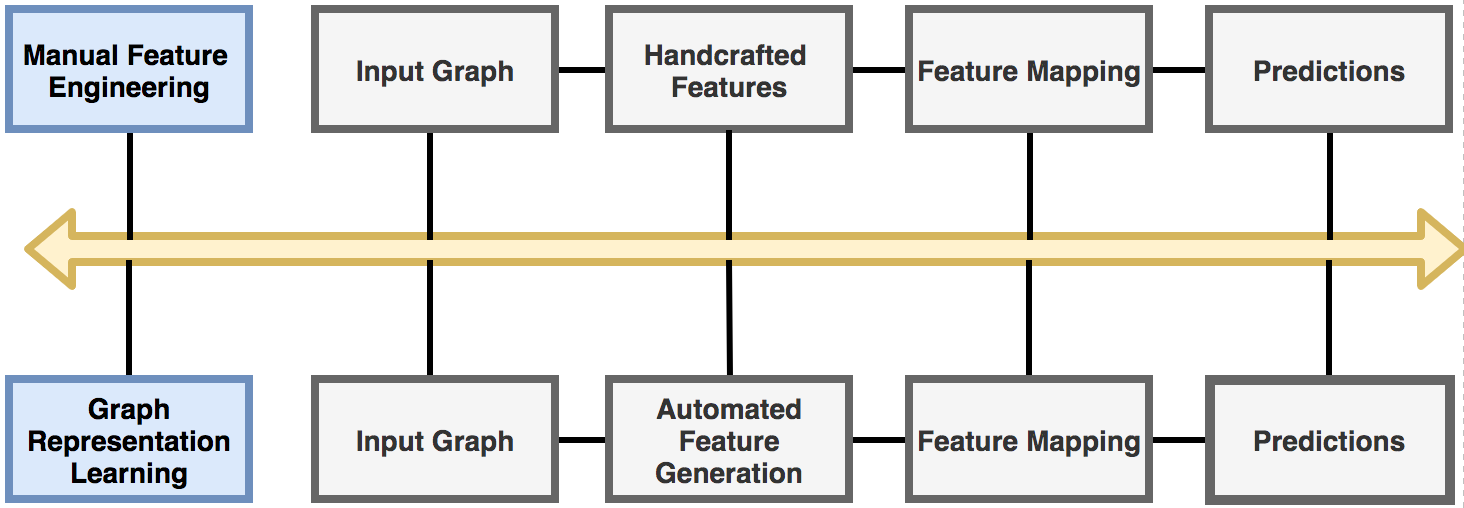}}
\caption{Manual Feature Engineering versus Graph Representation Learning}
\end{figure}

The graph representation or graph embedding based learning models learn and encode graph features with nodes into low dimensional space. It can be trained incorporating dimensionality reduction \cite{belkin2001laplacian} and neural networks. The utilization of GRL prompted the improvement of cutting edge language models like dialogue systems, relation prediction and  natural language understanding. Existing well-known and standard models such as CNN \cite{albawi2017understanding}, RNN \cite{mikolov2010recurrent} and Word2Vec \cite{goldberg2014word2vec} can be employed. However, these models are generally used for grids (in case of CNN) and sequences (in case of RNN). In addition, graphs are more complex due to no spatial locality, no fixed ordering and an isomorphism problem \cite{fortin1996graph} which is one of the most critical aspects when designing an architecture. Graphs therefore tend to be more connected and complex. It may also contain multimodal knowledge like nodes representing concepts, text, numbers and timestamps that further adds to the complexity. We need specific models tailored to graphs only. Due to this researchers came up with graph representation that is learning representation of nodes and edges to turn facts into vector representations. Feeding the graph to a vector space is called encoding. The reproduction of the graph neighborhood from the embeddings represents decoding. The vital advantage of the encoder-decoder structure is that it permits to concisely characterize and analyze diverse modeling strategies dependent on the similarity values, loss and decoder function \cite{hamilton2017representation}.

\subsubsection{Graph Feature Based Methods}

One method of reviewing the encoder decoder approach is according to the viewpoint of matrix factorization. Of course, the test of interpreting local neighborhood structure from vector space of nodes is firmly identified with rebuilding items in graph adjacency matrix.  We can see this approach as utilizing factorization of matrix to gain proficiency with a low-dimensional estimation for a node to node similarity matrix M, where S sums up the adjacency matrix and gains concepts of similarity between nodes. The fundamental reason for employing matrix factorization based techniques is to diminish the dimensionality while likewise protecting the locality and nonlinearity of the graph. Nonetheless, the global structural information may be lost. SVD (Singular Value Decomposition) is commonly used because of its realization in the low-position approximation \cite{abdi2007singular}.

Later works frequently employ decoders based on the inner product (e.g. graph factorization (GF) \cite{ahmed2013distributed} and HOPE \cite{ou2016asymmetric}) on the assumption that the closeness between two nodes (the cover between their nearby areas) is relative to the dot product of their embeddings for link prediction. The Graph factorization model works by limiting the quantity of adjoining nodes for cutting the graph, as opposed to applying edge cuts. HOPE is centered around modeling and representing directed graphs on the premise that directed relations can address any type of graph. For directed graph embeddings, HOPE also features the asymmetric and transitive properties. It additionally upholds traditional similarity based methods such as common neighbors (CN), academic adar index (AAI), katz index (KI) and rooted pagerank (RPR).

The objective of GF techniques is to learn embeddings for every node to such an extent that the inner product between the vectors of the learned embeddings approximates some deterministic proportion of the overlap between their local neighborhood areas. Late years have seen a flood in productive techniques that adjust the inner product technique to deal with utilization of stochastic proportions of node similarity. A vital advance in these methodologies is that the node embeddings are upgraded so that two nodes have comparable embeddings if they can coexist on a short random walk. DeepWalk \cite{perozzi2014deepwalk} and node2vec \cite{grover2016node2vec} employ a shallow embedding technique and an inner product decoder. The key difference in these techniques is how they portray the concepts of similarity between nodes and neighborhood remaking. Instead of directly remaking the adjacency matrix, they forward embeddings to encode the measure of the random walk. To generate random walk embeddings, the overall methodology is to use decoders and limit the cross entropy loss. However, easily estimating the cross entropy loss can be computationally expensive.

There are various systems to conquer this computational test and this is one of the fundamental contrasts between the DeepWalk and node2vec methods. DeepWalk utilizes a progressive softmax to inexact cross entropy loss which includes incorporating a binary tree design to speed up the calculations whereas node2vec employs a noise contrastive technique with the approximation of normalizing factor utilizing negative facts. The node2vec approach likewise separates itself from the prior DeepWalk calculation by considering a more adaptable meaning of random walks. To be specific, DeepWalk utilizes random walks consistently to characterize the operation of decoder whereas the node2vec algorithm presents hyper parameters that permit the random walk probabilities to easily add between walks that are more likened towards the BFS or DFS on the graph. A change to the node2vec variation, graph2vec basically focuses on how to properly embed a subgraph from a graph \cite{hamilton2020graph}. 

Dissimilar to the aforementioned strategies, SDNE (Structural deep network embedding) \cite{wang2016structural} does not utilize random walks. It attempts to gain information from two particular measurements namely first order proximity and second order proximity. In the former approach, two nodes are considered comparable in the event that they share an edge whereas in the latter one, they are considered related if they share many adjoining or nearby nodes. The objective is to catch deep non linear patterns. The first order proximity is preserved using a graph dimensionality reduction algorithm namely laplacian eigen maps \cite{belkin2001laplacian} whereas second order proximity is preserved by utilizing an unsupervised autoencoder that contains reconstruction loss function to minimize. Both loss functions are then jointly minimized to obtain graph embedding.

LINE \cite{tang2015line} also adopts the similar methodology by defining first and second order proximity. Its main functionality is to decrease the range of the distinction between the input and embedding distributions and is accomplished by utilizing KL divergence \cite{ponti2017decision}. It generates two probability distributions (adjacency matrix and dot product) for each pair of nodes and decreases the KL divergence. The main drawback of this approach is that it does not perform well overall if the application requires insights relating to node neighborhood structure. This is because it needs to mark new functions for each extended sequence of proximity.

\begin{figure}[htbp]	
	\centering
	\begin{subfigure}[t]{1in}
		\centering
		\includegraphics[width=1in]{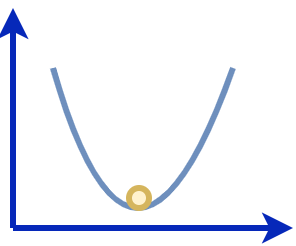}
		\caption{}\label{fig:1a}		
	\end{subfigure}
	\quad
	\begin{subfigure}[t]{1.2in}
		\centering
		\includegraphics[width=1.2in]{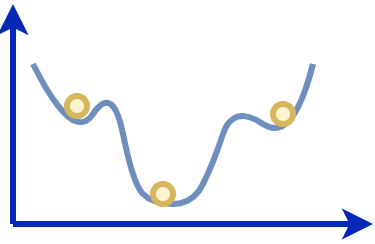}
		\caption{}\label{fig:1n}
	\end{subfigure}
	\caption{(a) Convex (b) Non Convex}\label{fig:1l}
\end{figure}

One method worth mentioning is HARP \cite{chen2018harp} which improves the performance of the walking and embedding based techniques mentioned above. Due to the non-convex based objective functions in the previous models, they can be trapped in local optima (as shown in figure 4) and therefore reduce performance. HARP on the other hand preprocess the graph using coarsening techniques to accumulate associated nodes into super nodes. In any case, it is important to understand that shallow embedding approaches experience the negative effects of some significant disadvantages. The first issue is that they do not share the parameters among nodes in the encoder, since it straightforwardly processes a unique embedding vector for every node. This absence of parameter sharing is computationally and statistically expensive.  A subsequent major drawback that questions the shallow embedding approaches is that they don't have the ability to utilize node features in the encoder. The shallow embedding based techniques can create embeddings for nodes that were available during the preparation stage only. Creating embeddings for new nodes (after training stage) is not possible except if extra processing steps are performed to gain proficiency with the embeddings for these nodes. These limitations forestalls shallow embeddings from being utilized on inductive applications \cite{hamilton2017representation}, which include summing up to hidden nodes subsequent to training. To mitigate these restrictions, shallow encoders can be substituted with more refined encoders like graph neural networks \cite{scarselli2008graph} that depend for the most part on the structure and properties present in the graph.

\subsubsection{Graph Neural Networks}

The overall idea of graph neural networks was introduced in \cite{scarselli2008graph}, however various neural network based models have been introduced for inferring multi relational representation. Learning from subgraph, entities and attributes (SEAL) \cite{zhang2018link}, exploit graph neural networks to learn structural and latent information in graphs for link prediction. It works by preparing the features from local enclosing subgraphs for each target link and feeding it into GNN to predict the missing links. HetGNN \cite{zhang2019heterogeneous} on the other hand takes into account heterogeneous networks and begins random walk with restart methodology and tests a fixed size of associated heterogeneous neighbors to cluster them dependent on type of nodes. In this way, the neural network design with two modules is used to aggregate the feature data of the adjacent vertices examined. The first module is responsible to inherit the content vector space for every vertex; on the other hand, the aggregate of content embeddings generated is completed by the second module. HetGNN then ensembles the generated outputs to acquire the ultimate embedding.  

The GNN's are further classified into types such as GAE (Graph AutoEncoder) \cite{han2020gaeat} and  VGAE (Variational GAE) \cite{kipf2016variational} (for example GCMC \cite{berg2017graph} and ARGA \cite{pan2018adversarially}) aim to leverage unsupervised learning to become familiar with node representations in a graph. GAE can train with the structural features in graphs while exploiting neural networks, and lessen the graph dimensionality as per the quantity of channels of the autoencoder covered up layers \cite{cao2016deep}. Moreover, GAE based models can install the entities to two dimensional vectors with assorted range. This advantages the auto-encoders exclusively to accomplish superior performance for evaluating over the obscured node embeddings, in addition to combining the characteristics of node to further enhance the predictive power \cite{hamilton2017representation}. 

\begin{figure*}[h]
\centerline{\includegraphics[width=14cm, height=4cm]{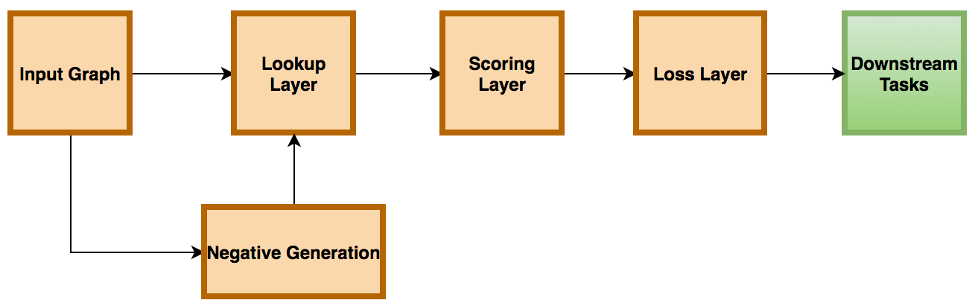}}
\caption{A bird’s eye view on knowledge graph embedding pipeline}
\end{figure*}

The representation procedures that depend on GNN consider both structural information and node attributes; however, they experience the effects of high complexity and shortcoming in recursive refreshing of the hidden states. Besides, GNN exploits identical boundaries for all layers, which restricts their adaptability. Taking advantage of convolutional neural networks (CNNs), GCNs lead to flexibly deriving features from complex graphs. The iterative combination of a node locality is leveraged by GCN to acquire graph embeddings. Here the accumulation technique prompts higher versatility other than learning global neighborhoods in graphs. Moreover, GCNs can also be used for subgraph embeddings \cite{chiang2019cluster}. 

GraphSAGE \cite{hamilton2017inductive} is one of the earliest models that aggregates the data from nearby neighborhoods iteratively. This recursive feature may generalize the model to hidden nodes. The nodes ascribed for this model may incorporate basic node measurements, like node degrees, literary information for profile data or for online social graphs. RGCN (Relational Graph Convolutional Networks) is introduced in \cite{schlichtkrull2018modeling} with the aim to predict the missing links for relational data types utilizing GCN.  This model is not quite the same as would be expected GCNs as the aggregation of features vectors of adjoining nodes are specific to relations. The errand of link expectation by this model can be seen as processing representation of nodes with a RGCN encoder and employing DistMult factorization \cite{yang2014embedding} as the scoring capacity. We talk about different GCN based strategies for KGC in Section 3.3.3.

\section{Knowledge Graph Embeddings}

KGE are models that attempt to learn the embeddings, i.e., vector representation of nodes and edges, by taking advantage of supervised learning. They do that by projecting entities and relationships into a continuous low dimensional space. These vectors have a few hundred dimensions which suggests memory efficiency. A vector space in which each point represents a concept and the position in the space of each point is semantically meaningful, similar to word embeddings. Examples of KGE models include RESCAL \cite{nickel2011three}, TransE \cite{bordes2013translating}, DisMult \cite{yang2014embedding}, ComplexE \cite{trouillon2016complex}, HolE \cite{nickel2016holographic}, ConvE \cite{dettmers2018convolutional}, RotatE \cite{sun2019rotate} and many more. All these competing models try to achieve a goal by learning a meaningful set of embeddings by maximizing the chance of predicting a test set of missing links. An ideal KGE model should be expressive enough to catch KG properties, for example, symmetric, asymmetric, inversion, and composition, that address the ability to represent distinctive logical patterns for relations \cite{sun2019rotate}. 

\begin{mydef}
Symmetric Relation: A relation R is symmetric if $\forall$ e1, e2: (e1, R, e2) $\rightarrow$ (e2, R, e1). 

For Example: e1 = John and e2 = Peter and R = "is brother of"; (e1, R, e2) = John is brother of Peter $\rightarrow$ (e2, R, e1) = Peter is brother of John.
\end{mydef}

\begin{mydef}
Asymmetric Relation: A relation R is antisymmetric if $\forall$ e1, e2: (e1, R, e2) $\rightarrow$ $\neg$(e2, R, e1)

For Example: e1 = John and e2 = Peter and R = "is a supervisor of"; (e1, R, e2) = John is a supervisor of Peter $\rightarrow$ (e2, $\neg$R, e1) = Peter is not a supervisor of John.
\end{mydef}

\begin{mydef}
Inverse Relation: A relation Ri inverse to relation Rj if $\forall$ e1, e2: Rj(e1, e2) $\rightarrow$ Ri(e2, e1). 

For Example: e1 = John, e2 = Peter, Ri = "is a supervisor of" and Rj = "is a student of"; (e1, Ri, e2) = John is a supervisor of Peter $\rightarrow$ (e2, Rj, e1)= Peter is a student of John.
\end{mydef}

\begin{mydef}
Composite Relation: A relation Ri is composed of relation Rj and relation Rk if $\forall$ e1, e2, e3 : (e1, Rj, e2) $\wedge$ (e2, Rk, e3)$\rightarrow$(e1, Ri, e3). 

For Example:  e1 = John, e2 = London, e3 = United Kingdom, Rj = "is born in", Rk = "is capital of", Ri = "is from"; (e1, Rj, e2) = John is born in London $\wedge$ (e2, ,Rk, e3) = London is capital of United Kingdom $\rightarrow$ (e1, Ri, e3)= John is from United Kingdom.
\end{mydef}

Considering KG properties helps differentiate the representation limits of KGE decoders. Although we cannot infer that these patterns should hold precisely in practice, there might be numerous relations that show these examples to a certain extent. For instance,  symmetric relations hold more than 90\% of the time. Table 1 reproduced from \cite{sun2019rotate} sums up the capacity of the different decoders to encode the aforementioned KG properties.  KGE models ought to likewise show hierarchies, type constraints, transitivity, homophily, and long range dependencies. A good embedding should model these properties as best as possible while keeping a good tradeoff upon expressivity, scalability, and time to train a model.

A bird's eye view on learning the multi relation embeddings comprises various stages, as given in figure 5. To start with, the embeddings of both the entities and relations are first introduced utilizing random noise. These generated embeddings are then used to assign a score for correct and incorrect facts by employing a scoring function that learns their interaction. The embeddings are then updated using the optimizer function to limit the training loss on the triples scored. During KGE training, we are basically learning how to place vectors in an embedding space. The main task is to provide a maximum score for correct facts and less score for negative facts.

\begin{table*}[htbp]
\caption{A Comparison of KGE models in terms of capturing relation types \cite{sun2019rotate}}
\begin{center}
\begin{tabular}{|c|c|c|c|c|}
\hline
\textbf{Model}& \textbf{Symmetry}& \textbf{AntiSymmetry}& \textbf{Inversion}
& \textbf{Composition}\\
\hline
SE&False&False&False&False  \\
TransE&False&True&True&True  \\
TransX&True&True&False&False  \\
DistMult&True&False&False&False  \\
CompleX&True&True&True&False  \\
RotatE&True&True&True&True  \\
\hline
\end{tabular}
\end{center}
\end{table*}

\subsection{Embedding Lookup Layer}
This layer is responsible for mapping the one hot encoding vector to embedding vectors. The one hot vector represents a discrete sparse vector addressing an input. For a triple (e1, R, e2), three one-hot encoding vectors are required to map e1, R, e2, respectively. The embedding vector, on the other hand, is a low dimensional space containing semantically meaningful associations. It reduces the sparsity and leads to productive distributed representations.

\subsection{Negatives Generation}
An important step in training the KGE model is negative generation, which researchers have not fully emphasized in recent years. However, attempts have been made to fully exploit and generate corruption to help rectify scalability issues. There are two general assumptions, i.e., closed world assumption (CWA) and open world assumption (OWA). In CWA, the absence of a fact means that it is necessarily false, whereas, in the case of OWA, the absence of fact does not indicate that the fact is false.  Knowledge graphs operate under the open-world assumptions that means that if we process knowledge bases such as DBpedia \cite{bizer2009dbpedia} does not have false facts in it. The task of link prediction requires training models with false facts that can separate the true facts from them. Therefore, the CWA \cite{dong2014knowledge} is used to assume that the KG is only locally complete. To avoid insignificant predictions from the embedding, a complete set that includes all the corrupted facts should be handcrafted. Then, while reflecting on calculation cost and memory space, stochastic preparation is required at each step. In particular, to train KGE whenever we get a positive fact, we need to test some corrupted triples from its related negative sampling set. When generating corruptions under the CWA assumptions, we always try to corrupt either the subject or the object, as given in the equation below:

\begin{equation}
    Corruptions = \{(s', p, o) | s' \epsilon E\} \cup \{(s, p, o') | o' \epsilon E\}
\end{equation}
Here s' and o' represents corrupted subject and object. It is worth noting to mention some negative sampling techniques \cite{qiannegative} such as uniform sampling \cite{bordes2013translating}, Bernoulli sampling \cite{wang2014knowledge}, KBGAN \cite{cai2017kbgan}, IGAN \cite{wang2018incorporating}, NSCaching \cite{zhang2019nscaching}. 

Uniform sampling  \cite{bordes2013translating} refers developing negative triples by substituting the tail or the head entity of a positive triplet with the element arbitrarily tested from the entity set by the uniform distribution. Nonetheless, this sampling technique develops directly classified triplets that do not contribute to providing meaningful and important knowledge \cite{sun2019rotate} \cite{zhang2019nscaching}. Thereafter, as the preparation progresses, a large proportion of the tested negative triples obtain very low scores and almost zero gradients, hindering the preparation of the embedding model after just a few cycles of recurrence. One more extreme downside of uniform sampling is generating facts that appear negative when they should not. Subsequent to supplanting the head in (KamalaHarris, Gender, Female) with NikkiHaley, (NikkiHaley, Gender, Female) is verified truth. To lighten this issue, Bernoulli sampling \cite{wang2014knowledge} was proposed substituting head or tail elements with various probabilities as per the mapping property of relationships. Nonetheless, for relations with less information, it neglects to foresee the missing facts among semantically potential choices even after many epochs of training. Probabilistic negative examining \cite{kanojia2017enhancing} speeds up the most common way of producing negative triples by acquiring a train bias tuning boundary that decides the likelihood by which the created negative facts are supplemented with early-recorded potential examples. To resolve the issue of simple negatives, self adversarial sampling was proposed \cite{sun2019rotate}, which gauges each inspected negative as per its likelihood beneath the embedding models. On the other hand, the literature \cite{cai2018kbgan} \cite{wang2018incorporating} proposed sampling techniques leveraging Generative Adversarial Networks (GANs) \cite{goodfellow2014generative} that are powerful and effective but expensive to formulate and require black-box evaluation methods, and are not interpretable. In contrast to the aforementioned GAN-based strategies, one rich methodology that uses fewer boundaries and is simple to prepare is NSCaching \cite{zhang2019nscaching}, which includes employing a cache or reserve of strong negative triples with high scores.

\subsection{Scoring layer}
The scoring layer interacts with the loss function. A scoring function ($f_{r}(h,t)$) assigns a score to a triple (s, p, o). The higher score represents a higher probability of the triplets being true facts. There are several ways to design a scoring function. Some functions determine a model that scales better than others, and some are designed to capture properties in KG such as symmetry, asymmetry, homophily, etc. The three main categories of scoring functions for the KGE model, i.e., translation, factorization, and neural networks, are discussed in the following subsections.

\subsubsection{Translation Models}

Since the advent of the word embedding model, word2vec \cite{goldberg2014word2vec}, a lot of progress has been made to embed the representation learning in a distributed manner. Researchers are attempting to explore the interesting translation invariance phenomenon generated by the word2vec model. In simple words, the vector space generated by word2vec contains intrinsic semantically meaningful examples that can help capture the properties of facts for a better representation space. For example, Man is semantically related to a Male, whereas Woman is semantically related to a Female. Inspired by the word2vec model, TransE \cite{bordes2013translating}, a translation-based knowledge graph embedding model is proposed to capture the translation invariance phenomenon in multi-relational graphs. The principle thought behind adopting this approach is to acknowledge the most general and interpretable way of discovering legitimate triples as the translation activity of elements, characterizing the scoring function, and afterward limit the loss function to become familiar with the embedding of triples. TransE is responsible for modeling the entities (e) and relations (r) in uniform low dimensional continuous space Rd. As shown in figure 6, assuming the triple (h, r, t) is valid, the t generated is near the vector representation of h and r. As seen, TransE follows a mathematical principle given below: 
\begin{equation}
h + r \approx t
\end{equation}

\begin{figure}[htbp]
\centerline{\includegraphics[width=3cm, height=2cm]{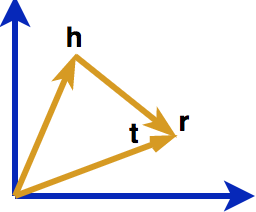}}
\caption{TransE \cite{bordes2013translating}}
\end{figure}

Here the triplet (h, r, t) consists of a head unit (h), tail unit (t), and the relation (r) between them, which are embedded in the vectors h, r, t. The scoring function of TransE is given in the equation below:
\begin{equation}
f_{r}(h,t) = ||\hspace{1mm} h + r - t \hspace{1mm}||\hspace{1mm}_{l1/l2}
\end{equation}
where  $l1/l2$  are the norm constraints. The TransE has repeatedly shown good performance for large scale knowledge graphs. Nonetheless, it fails to effectively model the complex relations such as one to many, many to many \cite{lin2015learning}. More specifically, assuming a one to many relation (i.e., for each head element, there are multiple tails elements associated with it) in which the ResearchArea depicts one to many relation between triplets (Ram, ResearchArea, ComputerVision) and (Ram, ResearchArea, LanguageProcessing).  The embedding vectors generated by TransE for ComputerVision and LanguageProcessing will be somewhat similar in the feature vector space. But this outcome is completely unreasonable on the grounds that ComputerVision and LanguageProcessing are entirely different fields. 

Curbing the limitation of the TransE model, Ma et al. proposed TransH \cite{wang2014knowledge} to give different representation vectors to each entity depending on the relation. In other words, TransH works by issuing an entirely separate relation-specific hyperplane for each relationship so that the entities associated with it have different semantics only in the context of that relationship. As shown in Figure 7, for the entity embedding vectors h and t, TransH projects it to the hyperplane (relation specific) in the direction of mapping vector $W_{r}$ that gives the projection vector $h_{\perp}$ and $t_{\perp}$.  

\begin{figure}[htbp]
\centerline{\includegraphics[width=4cm, height=2.5cm]{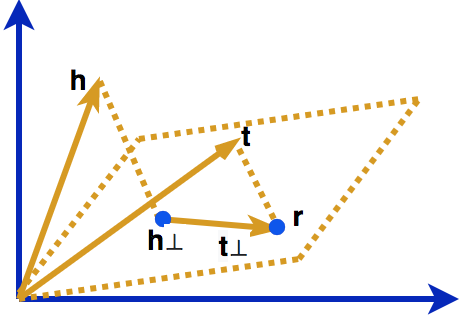}}
\caption{TransH \cite{wang2014knowledge}}
\end{figure}
The score function of TransH is formulated as follows: 
\begin{equation}
f_{r}(h,t) = ||\hspace{1mm} h_{\perp} + D_{r} - t_{\perp} \hspace{1mm}||
\end{equation}
Here $D_{r}$ represents relation specific translation vector, $h_{\perp}$ and $t_{\perp}$ follows the calculation approach given below: 
\begin{equation}
X_{\perp} = X - W^T_{r}XW_{r}
\end{equation}
where $W_r$ represents the normal vector of hyperplane that mentions $D_{r}$. TransH to some extent solves the problem related to complex relations by modeling each entity to different representation vectors dependent on the relation. However, it still employs the same vector feature space, $R_{d}$, for representing the facts. In general, an entity may have multiple semantics, and the relations are centralized towards numerous aspects of the entity.

TransR \cite{lin2015learning} attempts to model the entities utilizing relation specific vector space. The relations are modeled as a vector r particular to relation space $R_{s}$. As shown in figure 8, it operates by projecting the h and t from entity space ($R_{d}$) to relation specific space ($R_{s}$) generated by projected vectors $h_{\perp}$ and $t_{\perp}$.
\begin{figure}[htbp]
\centerline{\includegraphics[width=6cm, height=3cm]{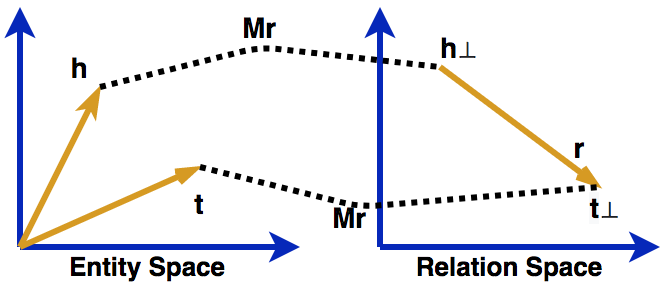}}
\caption{TransR \cite{lin2015learning}}
\end{figure}
The scoring function is similar to that given in equation 4 but with $h_{\perp}$ and $t_{\perp}$ as follows:
\begin{equation}
X_{\perp} = M_{r}X
\end{equation}
where $M_{r}$ is a projection matrix or mapping matrix generated by projecting entity vectors into relation specific space. Compared to TransE and TransR, TransR shows some competitive performance. However, it is also associated with limitations that need to be addressed. Without much trouble, one can understand that the semantics shared by the tail and head unit may be completely different. 
For example, triplet(John, research\_area\_is, computervision) in which the head entity John (person) is completely different from computervision (field of computer science). Nonetheless, in TransR, the projection matrix is the same for the head and tail unit for a particular relationship that directly impacts the predictive accuracy. It also suffers from high memory complexity because it creates a separate representation space for a relation that is not memory efficient.

TransD \cite{ji2015knowledge}, an improvement of TransR, adopts a dynamic mapping matrix that effectively generates two separate mapping matrices for head and tail entities. It exploits two embedding vectors for the representation of each entity and relation. The first embedding vector is used to represent the semantics of entity and relations (r belongs to $R_{s}$ and h, t belongs to $R_{d}$). The second embedding vector ($R_{m}$ belongs to $R_{s}$ and $H_{m}$, $T_{m}$ belongs to $R_{d}$) is employed to generate two dynamic projection matrices $(M_{h}, M_{t})$ as shown in figure 9.
\begin{figure}[htbp]
\centerline{\includegraphics[width=7cm, height=3cm]{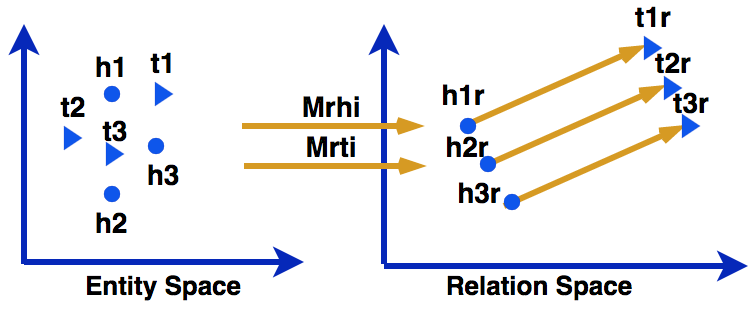}}
\caption{TransD \cite{ji2015knowledge}}
\end{figure}
The scoring function is given in equation 4 with $h_{\perp}$ and $t_{\perp}$ follows: 
\begin{equation}
X_{\perp} = M_{x}X
\end{equation}
\begin{equation}
M_{x} = R_{m}X^{T}_{m} + I_{m}
\end{equation}
where,  $I_{m}$ represents the Identity matrix. TransD replaces the vector product and matrix operations with the vector product operation in the previous model. This improves the computation effectiveness marginally and addresses the excessive number of hyperparameters in the TransR model, making TransD reasonable for huge scope KGs. 

Most of the aforementioned approaches failed to represent specific types of properties such as imbalance and heterogeneity. Heterogeneity indicates that some relations may have many connections to simple relations, which cause overfitting or underfitting (when the relationship is complex). On the other hand, the imbalance suggests treating the head and tail differently since there may be a high contrast present between them.  To handle these issues, a TranSparse \cite{ji2016knowledge} embedding method is proposed which is divided into two parts namely share and separate version.

In TranSparse (share) method, the projection metrics are substituted by an adaptive sparse matrix $M_{r}(deg_{r})$. The idea is to replace the dense features with sparse features to tackle the non-uniform assortment of relations and entities and decrease the number of hyperparameters in the model simultaneously \cite{ji2016knowledge}. The $deg_{r}$ represents the sparse degree dependent on the number of entities that are associated with relation r. Given scoring function in equation 4, the $h_{\perp}$ and $t_{\perp}$ follow: 
\begin{equation}
X_{\perp} = M_{r}(deg_{r})X
\end{equation}
\begin{equation}
deg_{r} = 1-(1-deg_{min})n_{r}/n_{r^{'}}
\end{equation}
Here, $deg_{min}$ is a hyper parameter that lies between 0 and 1. And, $n_{r}$ represent the number of entities that are associated with relation with  $n_{r^{'}}$ constitute the maximum of them. In TranSparse (separate) method, two sparse mappings are employed for projecting the head  $(M_{rH}(deg_{rH}))$ and tail $(M_{rT}(deg_{rT}))$ separately for each relation. Given scoring function in equation 4, the $h_{\perp}$ and $t_{\perp}$ follow: 
\begin{equation}
X_{\perp} = M_{rX}(deg_{rX})X
\end{equation}
\begin{equation}
deg_{rX} = 1-(1-deg_{min})n_{rX}/n_{rX^{'}}
\end{equation}

To simplify the execution of the TranSparse embedding, the literature propose sTransE \cite{nguyen2016stranse} that works by simply replacing the projection sparse matrix by mapping matrix. The projected vectors are extended as follows:
\begin{equation}
X_{\perp} = M_{rX}X
\end{equation}

The aforementioned approaches merely focus on modifying projection vectors, mapping matrices, and embedding spaces. However, none of the techniques were taken advantage of in employing better optimization techniques to increase the predictive power of the standard TransE model. The TransA \cite{xiao2015transa} model optimizes the TransE model by replacing the distance measure from the standard Euclidean distance to adaptive Mahalanobis distance as it provides more flexibility and adaptability managing complex relations. Given a nonnegative symmetric weighted matrix $M_{r}$ (with relation r), the scoring function of TransA is defined as follows:
\begin{equation}
f_{r}(h,t) = (|h+r-t|)^{T}M_{r}(|h+r-t|)
\end{equation}

Other than permitting entities to possess different embedding when engaged with various relations, a different line of exploration augments TransE by weakening the overstrict prerequisite given in equation (2). The equation (15) represents the scoring function of TransM \cite{fan2014transition}. It maps each triplet with weight particular to a relation $\Theta_{r}$. By allocating low weight to complex relations (i.e., one to many, many to many) it permits the tail entity to place afar from $h + r$.
\begin{equation}
f_{r}(h,t) = -\Theta_{r}||h+r-t||_{1/2}
\end{equation}
ManifoldE \cite{xiao2015one} employs the concept of hypersphere by relaxing equation(2) with $\Theta^{2}_{r}$ for each triplet belonging to the set of all facts. With this, the tail entity can lie roughly on a hypersphere with a diameter of 2$\Theta_{r}$ focused at h + r, instead of near the specific place of h + r. The score capacity is henceforth outlined as: 
\begin{equation}
f_{r}(h,t) = -(||h+r-t||^{2}_{2}-\Theta^{2}_{r})^2
\end{equation}
TransF \cite{feng2016knowledge} adopts a similar approach. Rather than upholding the precise interpretation given in equation(2), it expects the tail entity to place in the same direction of $h + r$, like $h$ with $t$$-$$r$. The scoring function has to coordinate  $h+r$ with $t$ as well as $t$$-$$r$ with $h$,
\begin{equation}
f_{r}(h,t) = (h+r)^{T}t+(t-r)^{T}h
\end{equation}

Recently, Xie et al. \cite{xie2017interpretable} proposed ITransF to reduce the data sparsity problem shown by TransE and STransE. They used the concept of sparse attention mechanism responsible for locating hidden concepts with statistical strength transfer through concept sharing. Besides, the learned relationship among concepts and relations, addressed by sparse attention vectors, is interpretable. The scoring function is given by: 
\begin{equation}
f_{r}(h,t) = ||\alpha^{H}_{r}.D.h+r-\alpha^{T}_{r}.D.t||_{l}
\end{equation}
Here, $D$ represents a concept projection matrix composed by normalized attention vectors $\alpha^{H}_{r}$, $\alpha^{T}_{r}$ belongs to $[0, 1]^{m}$ adopting convex combinations. Qian et al. \cite{qian2018translating} recommend that previous models fail to attract attention by disregarding the hierarchical structure of characteristics of the entities of human cognition and proposed TransAt \cite{qian2018translating}. It coordinates translation embedding utilizing an attention mechanism. To deal with more complex relations, literature proposed TransMS \cite{yang2019transms} which uses the concept of multidimensional semantics. It captures the semantics for the relation to the head or tail entity and from the head or tail entity to relations and between the entities employing nonlinear $tanh$ function.

A large part of current techniques is centered around the structured knowledge of triplets and augments the chance of their foundation \cite{wang2017knowledge}. However, they overlook the semantic data and the earlier information shown by semantic data incorporated in most KG. Taking advantage of the semantic encoding of data, TransT \cite{ma2017transt} was proposed for structured data representing the range of an entity and coordinate entity type. The entity type is responsible for creating the relationship type. Semantic analogy based on related elements and types of relationships is used to capture the prior arrangement of relations and entities. The models presented so far translate triples as deterministic focuses in vector space. New works consider Gaussian embeddings to counter the uncertainty and translate it as random variable \cite{he2015learning} \cite{xiao2015transg}. KG2E \cite{he2015learning} sees relations and entities as vectors drawn randomly from multivariate Gaussian distribution $(G_{d})$ and scores a triple utilizing the distance between the two arbitrary vectors. The TransG \cite{xiao2015transg} model additionally translates entities with G, utilizing a combination of $G_{d}$ to acquire various semantics. These models consider the uncertainty of the facts; however, this outcome is a complex model.

Techniques, for example, QuatE \cite{zhang2019quaternion}, RotatE \cite{sun2019rotate} and TorusE \cite{ebisu2018toruse} exploits quaternions, rotations, lie groups respectively and are like TransE. They do not supersede distance-based models fundamentally. However, their thought is equivalent to that of translation-based embeddings. Given a fact or triplet $(h, r, t)$, they all guide the head element to the tail element through the relationship $r$; however, the particular mapping function on $r$ is unique. Therefore this work places them in this subsection. RotatE \cite{sun2019rotate} develops a rotational hadmard product or element-wise multiplication ($\circ$) based on Euler's identity $e^{i\phi} =\cos\phi + i\sin\phi$. The relation is categorized as a rotation between the head and tail entity in a complex-valued space. The score function is formulated below:
\begin{equation}
f_{r}(h,t) = || h \circ r -t||
\end{equation}
RotatE also introduces a novel self-adversarial sampling technique that helps to effectively interpret the relationship types shown in Table 1. QuatE \cite{zhang2019quaternion} diversifies the complex space into 4-Dimensional hypercomplex space and uses a Hamilton product ($\bigot$) and acquires more meaningful rotational ability than RotatE. The scoring function of QuatE is represented below:
\begin{equation}
f_{r}(h,t) = ||h \bigot \frac{r}{|r|}.t||
\end{equation}
Even though the TransE can viably catch the properties in a KG by keeping a basic principle of $h + r \approx t$. It presents an issue of regularization by compelling embeddings of elements on a sphere in the vector space, which unfavorably influences the exhibition of the downstream tasks. TorusE \cite{ebisu2018toruse} tackles the above mentioned issue by adopting a Lie Group representing a n-dimensional torus space defined as $ \pi : R_{n} -> T_{n}, x -> [x]$ where $[h]$, $[r]$, $[t]$ belongs to $T_{n}$. Like TransE, it additionally learns embeddings following the translation in torus space:
\begin{equation}
[h] + [r] \approx [t]
\end{equation}

To sum up, KGC techniques dependent on the translation embeddings emphasized the utilization of the relations between entities, semantics between the relations and elements, and the structural data of the KG, which compensates for the complicated preparation and perturbing augmentation of conventional strategies. These techniques are fundamental and clear with solid interpretability \cite{miller1995wordnet}.

\subsubsection{Tensor Factorization Models}

This class of models classifies the function of the KGC task as a tensor decomposition belonging to the family of factorization models. It addresses the graph as a three-sided tensor that is disintegrated into a composition of low-dimensional element vectors. The principle thought is to ensure that the model does not overfit by employing a small number of common hyperparameters, making them easier and simpler to train. 

As shown in figure 10, it works by first creating a 3D binary tensor $\mathbb{X}$ (that is, $\mathbb{X}$ belongs to $R^{i.i.j}$ where $i$ and $j$ denotes the number entity and relation, respectively) utilizing the triplets present in multi-relational graphs. Each slice $\mathbb{X}_{s}$  where $s$ belongs to {0, 1, 2, to n} present in tensor $\mathbb{X}$  directly represents a relation type $R_{s}$. The value of $\mathbb{X}_{pqr}$ = 1 tells that $p^{th}$, $q^{th}$ entity and $r^{th}$ relation is present in the graph is true; otherwise, it indicates an undefined fact if the value equals zero.  

\begin{figure}[htbp]
\centerline{\includegraphics[width=4cm, height=3cm]{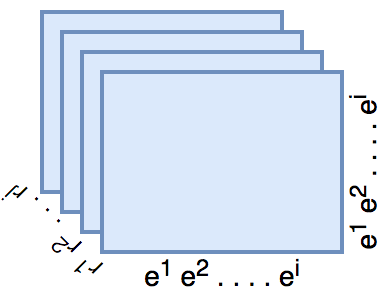}}
\caption{Knowledge graph representation as Tensors.}
\end{figure}

Rescal \cite{nickel2011three} is one of the earliest model to exploit this approach for capturing semantics. As presented in figure 11, it uses the rank-r factorization technique to capture the latent meaningful representation of the required structure present in the knowledge graph as a result of applying the tensor. For $s^{th}$ relation from the set of $m$ relations, equation (22) reflects tensor factorization in $s^{th}$ slice of $\mathbb{X}$:  
\begin{equation}
\mathbb{X}_{s} = AR_{s}A^{T}
\end {equation}
Here, $A$ represents an adjacency matrix responsible for capturing the latent semantic representation of entities. The pairwise interaction in $s^{th}$ relation is represented by matrix $R_{s}$. Given, $M_{r}$ is relationship matrix, the scoring function is defined as follows:
\begin{equation}
f_{r}(h,t) = h^{T}M_{r}t
\end{equation}

\begin{figure}[htbp]
\centerline{\includegraphics[width=6cm, height=3.5cm]{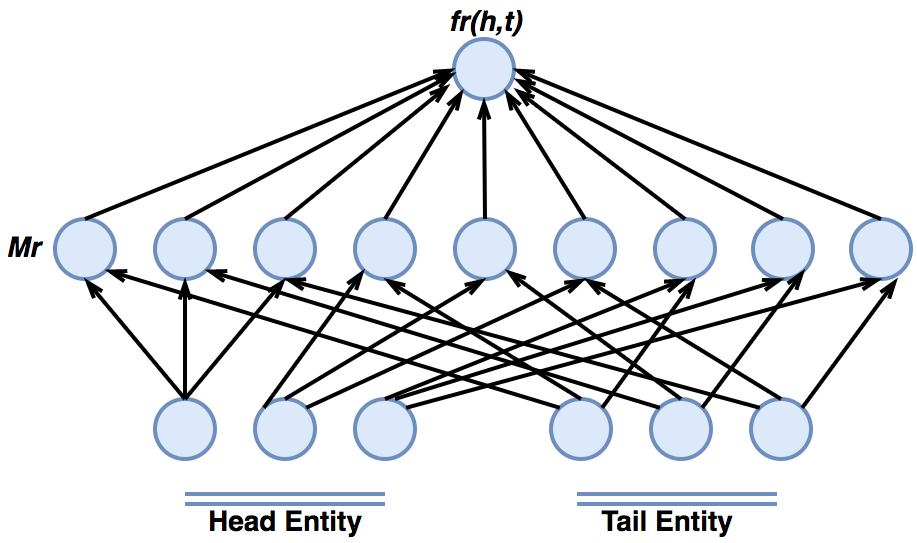}}
\caption{RESCAL \cite{nickel2011three}.}
\end{figure}

Fan et al. proposed TATEC \cite{garcia2014effective}, which is a more complex version of RESCAL by combining the two-way interactions between the entities and relations. To reduce the computationally complex nature of RESCAL, the DistMult \cite{yang2014embedding} suggests to incorporate only the diagonal matrix $dig(r)$ in place of $M_{r}$. The scoring function is then reduced to:
\begin{equation}
f_{r}(h,t) = h^{T}dig(r)t
\end{equation}
It infers the fundamental relations between sets of elements that are present in the same dimension. Contrasted with RESCAL model, it decreases the parameter count and remarkably augments the performance for extracting the target knowledge in graphs.
\begin{figure}[htbp]
\centerline{\includegraphics[width=4cm, height=3.5cm]{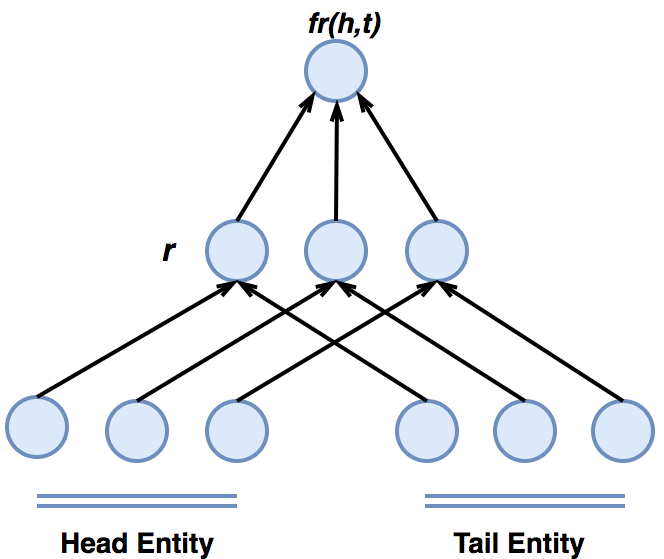}}
\caption{DistMult \cite{yang2014embedding}.}
\end{figure}

To capture the pairwise compositional properties between entities, HolE \cite{nickel2016holographic} introduces circular correlation operation \cite{plate1995holographic} denoted by $\star: R^{d} \times R^{d} \rightarrow R^{d}$. The scoring function of HolE is represented as:
\begin{equation}
f_{r}(h,t) = r^{T}(h \star t)
\end{equation}
Here, $\star$  is expessed as:
\begin{equation}
[a \star b]_{k} = \sum^{d-1}_{i=0}a_{i}b_{(k+i) mod(d)}
\end{equation}
The main intention behind adopting $\star$ is to leverage the reduced complexity of composite representation in the form of compressed tensor product. Furthermore, HolE makes use of the fast Fourier transform $f(.)$ \cite{stratos2010fast} which can further accelerate the computational process via:
\begin{equation}
a \star b = f^{-1}(\overline{f(a)} \circ f(b))
\end{equation}
A significant disadvantage of modeling with a circular correlation coefficient is that it is not composite. In simple terms, HolE cannot model asymmetric relations. Recently, Xu et al. \cite{xue2018expanding} proposed HolEX, which is Extended Holographic Embedding that interpolates both the full tensor product and HolE. Given $c$, a fixed vector belongs to $R^{d}$, for $a, b$ belongs to $R^{d}$, they introduced a perturbed holographic compositional operation which is defined as follows: 
\begin{equation}
h(a,b;c) = (c \circ a) \star b
\end{equation}

Most of the previous models exploit 3-way bivariate tensor decomposition for KGC. However, this methodology is not recommended for effectively capturing asymmetric relationships. Trouillon et al. \cite{trouillon2016complex} introduced the concept of leveraging the complex space $C^{d}$ to embed the entities and relations present in KG and proposed CompleX, which is a continuation of DistMult.  The CompleX can effectively infer asymmetric relationships. Instead of using a real-valued space, CompleX leverages complex space $C^{d}$ to embed the entities and relations. The scoring function is defined below: 
\begin{equation}
f_{r}(h,t) = Real(h^{T}dig(r)\overline{t}) = Real\bigg\{\sum^{d-1}_{k=0}(h)_{k}(r)_{k}(\overline{t})_{k}\bigg\}
\end{equation}
where, $\overline{t}$ represents the complex conjugate of the tail entity and $Real(.)$ denotes the real part of a complex relation.  Hayashi et al. \cite{hayashi2017equivalence} analyzed and studied the equivalence of Complex and HolE. It has been shown that the HolE is understood by CompleX as an exceptional example where conjugate symmetry is inflicted on the embedding, and alternatively, each complex has a corresponding HolE. 

Another interesting area of research is to integrate other reasoning regimes into knowledge graph embedding architectures. For example, if the $sun$ is surrounded by $planets$ and attracts $mass$ by analogical reasoning, then scale $sun$ to the $nucleus$ and $planets$ to the $electrons$. It can easily be concluded that the $nucleus$ attracts a $charge$ by analogy with the $sun$ attracting a $mass$. To infer the analogical reasoning in knowledge graphs, ANALOGY \cite{liu2017analogical} an extended version of RESCAL was proposed to model the characteristics of relations and entities as analogical properties by employing a bilinear scoring function as given in equation (23).
This equation however is followed by two major constraints that depend on the analogical properties given in equation (30, 31). First, it should be a normal matrix. Secondly, for each pair of relations, their structure of the linear map (that is, $M_{r}$) must be mutually commutative.
\begin{equation}
M_{r}M^T_{r} = M^T_{r}M_{r}
\end{equation}
\begin{equation}
M_{r}M_{r'} = M_{r'}M_{r}
\end{equation}

To address the issue of independence in taking advantage of the earliest tensor decomposition, a.k.a Canonical Polyadic (CP), Kazemi et al. \cite {kazemi2018simple} introduced SimplE, an extension of CP, to dependently learn two embeddings of each entity to simplify link prediction tasks. It introduces the inverse of relationships $(r')$ and calculates the mean score with scoring function via: 
\begin{equation}
f_{r}(h,t) = \frac{1}{2}\big{(}t\circ r't+h\circ rt \big{)}
\end{equation}

It has been shown that the earlier works have not focused on the utilization of crossover interactions, a.k.a bidirectional interactions between relations and entities that help to segment the related knowledge for KGC tasks \cite{zhang2019interaction}. The concept of related knowledge or information is explained in figure 13. 
\begin{figure}[htbp]
\centerline{\includegraphics[width=7cm, height=3cm]{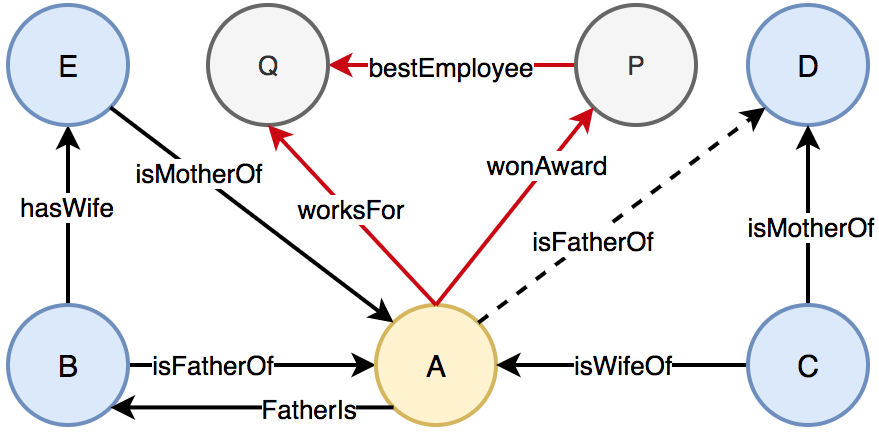}}
\caption{An example of made up Knowledge Graph \cite{zhang2019interaction}. The black-colored relationships represents the potential related knowledge and interaction to infer the dotted relation i.e. father-child relationship. The red-colored links depicting the professional relations of A do not give important knowledge to this undertaking.}
\end{figure}
The CrossE \cite{zhang2019interaction} model was proposed to exploit crossover interactions with an interaction matrix $C$ to obtain relationship specific embeddings $C_{r} = x^{T}_{r}C$. It uses hadmard product operation to incorporate head entity and relation with $C_{r}$ via:
\begin{equation}
h_{I} = C_{r} \circ h
\end{equation}
and,
\begin{equation}
r_{I} = h_{I} \circ r
\end{equation}
Here, $h_{I}$ and $r_{I}$ represents head interaction and relation interaction respectively. The score energy function of CrossE is then formulated as follows: 
\begin{equation}
f_{r}(h,t) = \sigma(\tanh(h_{I} + r_{I} + b)t^{T})
\end{equation}
where, $\sigma(x)$ is nonlinear function and $b$ represents bias vector.

\subsubsection{Deeper scoring function}

Over the years, deep learning has proven its implication in all specializations like computer vision, natural language processing, and graph-based learning. Researchers are attempting to take advantage of this cutting-edge technology in knowledge graph embedding to model complex nonlinear projections in continuous low dimensional space \cite{chen2020review}.

SME \cite{bordes2014semantic} or semantic matching energy characterizes energy functions employing neural networks, which can be utilized to quantify the certainty of each noticed reality $ \langle (s, p, o) \rangle$. As displayed in figure 14, initially each triplet is inserted to the vector space. Then, at that point, two projection matrices are applied to catch the semantic associations among elements and relations. The fully connected layer is then applied to finally calculate the semantic matching energy for each fact.  The main advantage of SME is that the relation type is not represented by a matrix, but is addressed by a vector. Therefore, in a situation where there are a large number of relation types, information related to the position of parameters and elements can be easily shared. The SME has two variants namely, bilinear form and linear form. In the two versions, the linear requires more computation time than the bilinear as it takes extra $n$ parameters for training. Given, $M_{x1}$, $M_{x2}$, $M_{y1}$ and $M_{y2}$ are projection matrices, the bilinear and linear form are represented in equation (36) and (37) respectively. 
\begin{equation}
\label{eq:m}
\begin{aligned}
g_{x}(h, r) = M_{x1}h + M_{x2}r + b_{x}\\
g_{y}(t, r) = M_{y1}t + M_{y2}r + b_{y}
\end{aligned}
\end{equation}
\begin{equation}
\label{eq:n}
\begin{aligned}
g_{x}(h, r) = (M_{x1}h) \circ (M_{x2}r) + b_{x}\\
g_{y}(t, r) = (M_{y1}t) \circ (M_{y2}r) + b_{y}
\end{aligned}
\end{equation}
Here, $b$ represents a global bias vector. The final energy score is achieved by combining the $g\_{x}(h, r)$ and $g\_{y}(t, r)$ as follows:
\begin{equation}
f_{r}(h, t) = g_{x}(h, r)^{T}g_{y}(t, r)
\end{equation}

\begin{figure}[htbp]
\centerline{\includegraphics[width=6cm, height=4.5cm]{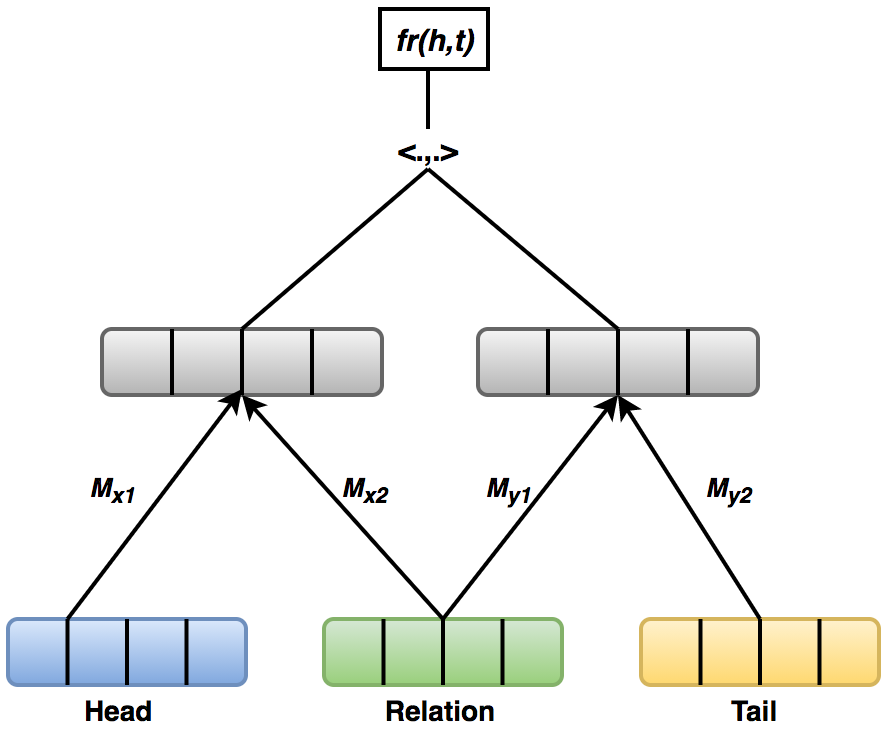}}
\caption{SME \cite{bordes2014semantic}.}
\end{figure}

The NTN or neural tensor network model \cite {socher2013reasoning} aims to substitute the standard linear layer in conventional neural networks with the bilinear tensor, and associate the head and tail element vectors in various arrangements as given in figure 15. Firstly, the elements in the triplet (that is, head and tail entity) are fed to the first layer which is responsible for mapping the given entities to the projection matrices $M_{r1}$ and $M_{r2}$ with the relation specific tensor $\chi_{r}$ in the second layer. Secondly, these three elements are then pushed to the third layer i.e. non-linear layer accountable for merging the features to obtain the semantic knowledge. The semantic information is then finally fed to the relation specific output layer to obtain the final score. Given, $g(x) = \tanh(x)$ and $b_{r} = bias$, the energy score of NTN is defined as follows:
\begin{equation}
f_{r}(h, t) =  r^{T}g(h^{T}\chi_{r}t + M_{r1}h + M_{r2}t + b_{r})
\end{equation}
The NTN can accomplish good accuracy for predicting obscure relations between entities. The presentation of tensors can precisely depict the complex semantic connection among elements. Although, there are problems with high computation cost due to the additional parameters introduced by relation-specific tensors that cannot be adjusted to represent KG on a large scale. In the same literature \cite {socher2013reasoning}, a less complex single layer model (SLM) model was proposed in which the value of relation specific tensor is null. The scoring function is obtained putting the value of $\chi_{r} = 0$ in equation (39): 
\begin{equation}
f_{r}(h, t) =  r^{T}g(M_{r1}h + M_{r2}t + b_{r})
\end{equation}
\begin{figure}[htbp]
\centerline{\includegraphics[width=6cm, height=4.5cm]{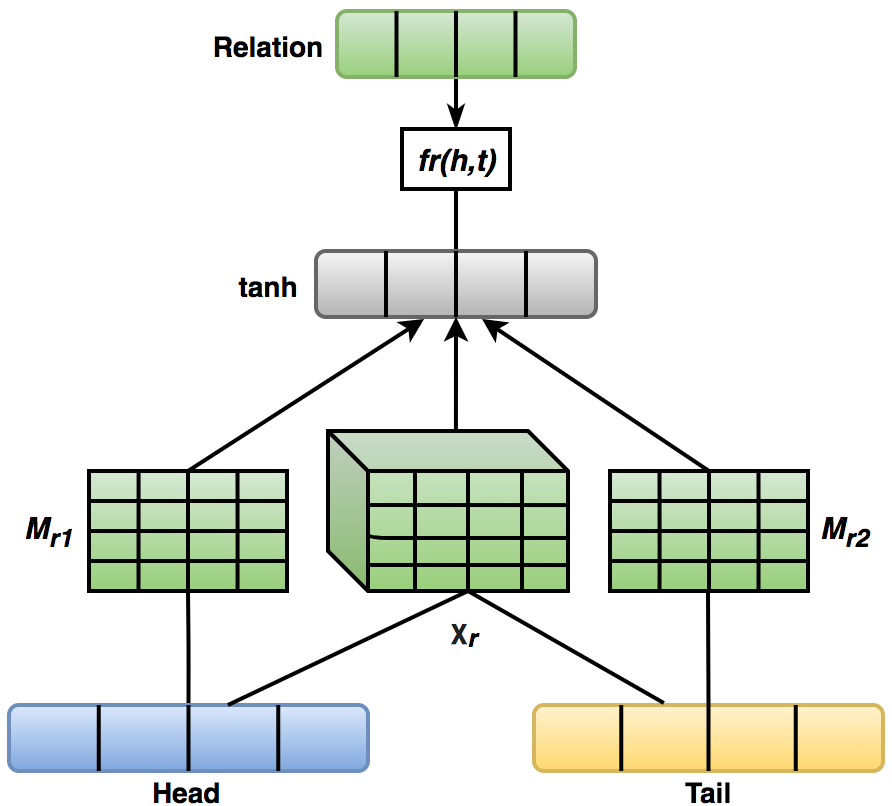}}
\caption{NTN \cite{socher2013reasoning}.}
\end{figure}

Curbing the limitations of NTN, Dong et al. introduced MLP \cite{dong2014knowledge} that gives a lightweight design for modeling the knowledge. It keeps all the elements i.e head entity, tail entity and relation at the same level that are simultaneously combined and projected into the vector space in the input layer. The generated matrices $(M_{i}, M_{j}, M_{k})$ are then fed into the nonlinear hidden layer $(\tanh)$ to generate the output score. The scoring function is defined as follows:
\begin{equation}
f_{r}(h, t) =  m^{T}g(M_{i}h + M_{j}r+ M_{k}t)
\end{equation}
\begin{figure}[htbp]
\centerline{\includegraphics[width=6cm, height=4.5cm]{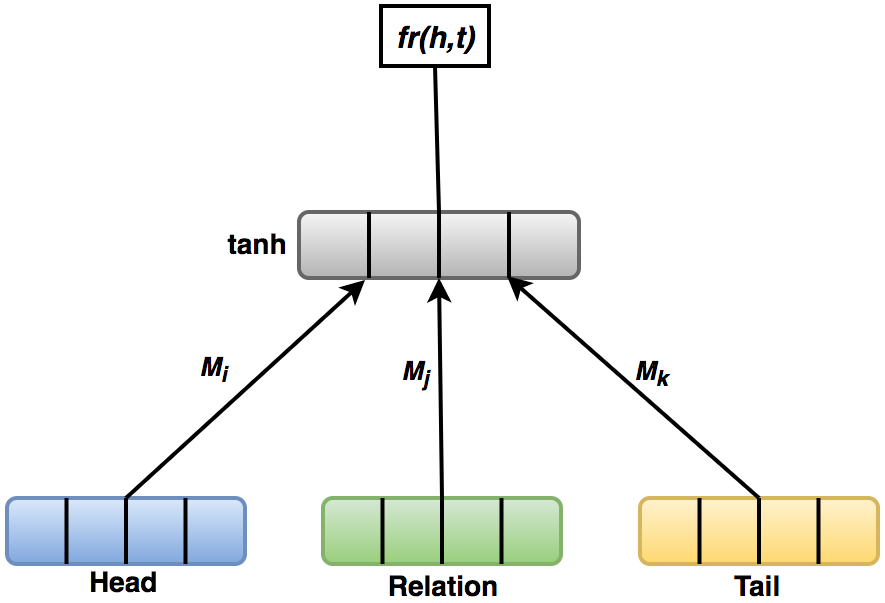}}
\caption{MLP \cite{dong2014knowledge}.}
\end{figure}

NAM or Neural Association Model \cite{liu2016probabilistic} sets up a deep neural architecture. The NAM embeds a given triplet into a feature vector space. At that point, the vector embeddings of the relation as well as the head entity are combined to get a vector $z_{0} = [h; r]$ which act as an input to architecture consisting of $N$ rectified units $(RELU)$ as follows:
\begin{equation}
\label{eq:t}
\begin{aligned}
a^{n} = M^{n}Z^{n-1} + b^{n}, \, \,\,\,\, where,  \, \, n = 1, 2, ...., N,\\
z^{n} =  RELU(a^{n}), \,\,\, \, \, where, \,\, n = 1, 2, ...., N
\end{aligned}
\end{equation}
The final score is generated by matching the output of the last hidden layer with the tail entity vector via:
\begin{equation}
f_{r}(h, t) =  t^{T}z^{N}
\end{equation}

ConvE \cite{dettmers2018convolutional} is one of the first models to use Convolutional Neural Networks (CNN) to predict missing links in knowledge graphs. Unlike fully connected dense layers, CNNs can help capture complex nonlinear relationships by learning with very few parameters. ConvE adopts embedded 2-dimensional convolution, which outperforms 1-dimensional convolution to capture the interactions between features for two embeddings. ConvE achieves local linkages between different entities in multiple dimensions; however, it ignores global relations of triple embeddings. First, it works by combining the head entities matrix and the relations it feeds into a 2-dimensional convolution to generate a feature map tensor. This tensor then goes through a linear transformation parameterized by matrix $M$ for projecting it into a low dimensional feature space. Finally, an inner product is used to match the tail unit. The scoring function for ConvE is defined as follows: \begin{equation}
f_{r}(h, t) =  g(vec(g(concat(\overline{e_{s}}, \overline{e_{r}})*w))M)e_{o}
\end{equation}
where, $concat$ is concatenation operator, $*$ represents convolution, and $e_{s}$ and $e_{r}$ are responsible for the 2D reshaping of the subject unit and relation unit, respectively.

Removing the reshaping activity from ConvE, ConvKB \cite{nguyen2017novel} uses 1D convolution to retain the interpretation properties of TransE, sufficient to capture global relationships and temporal attributes between entities. It addresses the embedding of each triple as a three-segment network and feeds it into the convolutional layer with the aim of achieving global connections between dimensional classes of a fact. The scoring potential of ConvKB is designed as follows: 
\begin{equation}
concat(g([e_{s}, e_{r}, e_{o}]*\Omega))w
\end{equation}
where,  $\Omega$ (filter set) and $w$ (weight vector) represents shared parameters. In HypER \cite{balavzevic2019hypernetwork}, the vector embeddings of each relationship are completely reshaped after projecting them through the dense layer, and subsequently, a bunch of convolutional channel weight vector relationships in each layer are adjusted. In contrast to linear combinations in ConvE, the non-linear and quadratic mix of element and connection embeddings gives HypER a much higher expressive range and the benefit of fewer hyperparameters. HypER can likewise be viewed as a factorization model. 

Capsule networks (CapsNets) \cite{sabour2017dynamic} is a new type of architecture introduced recently to limit the constraints of CNNs. The main problem with CNNs is that they are unable to predict translation-invariant events. For example, consider the task of guessing whether a cat is present in the given image. CNN can easily predict that the given image is of a cat. But, being unable to predict any additional information such as a change in the position of a cat. Nonetheless, CapsNets consists of many capsules and can capture the translation-invariant properties. A capsule is a small group of neurons where each neuron in the capsule represents different properties of a particular part of the given input followed by a dynamic routing process. 

Nguyen et al. \cite{vu2019capsule} adopted this approach and proposed CapsE to explore the state-of-the-art application of CapsNets in triple-based data. CapsE embeds a triplet $\langle h, r, t \rangle$ as unique d-dimensional vectors $V_{h}$, $V_{r}$ and $V_{t}$ respectively. The vectorized triplet $\langle V_{h}, V_{r}, V_{t} \rangle$ of $\langle h, r, t \rangle$ put into the convolution layer where different channels of a similar $1\times3$ shape are iteratively executed to generate d-dimensional feature maps which are then fed into neurons called capsules. As a result, each capsule can encode multiple features in setting up the triplet to address sections on the equivalent dimension, which are then sent to another layer containing a capsule to output the triplet score.

More recently, graph neural networks (GNNs) \cite{scarselli2008graph} have attracted much attention due to their incredible ability to represent graph structure. R-GCN \cite{schlichtkrull2018modeling} was perhaps the initial attempt to exploit GNNs for KGC. RGCN produces region-linked embeddings, feeds them to a decoder that predicts missing relationships in KG. Ordinary GCN cannot embed multi-relational graphs because it ignores edge knowledge present in the graph. Therefore, R-GCN somewhat replaces the scoring ability of basic GCN to capture the relationship between edges. However, R-GCN learns additional weight networks for every connection, consequently making the proposed work non-adaptable for enormous graphs. The authors attempted to quantify this issue with decomposition techniques, namely block diagonalization and basis \cite{zhang2018link}. These methods help to manage to overfit as well as make the connection weight matrices interdependent. In any case, the differential weight of the nodes locality is still undocumented, and a decoder is needed because they do not learn to embed relationships via graph neural nets. 

SACN \cite{shang2019end} attempts to augment the RGCN by adopting WGCN (Weighted Graph Convolution Network).  It accumulates data from the locality of nodes by being sensitive to edge relationship types. In WGCN, the entire graph is broken into subgraphs to such an extent that each subgraph contains edges of just a single connection type. Like the methodology in R-GCN, SACN uses WGCN as an encoder to comprehend entity embeddings, which is then fed to a decoder (Conv-TransE).  The aforementioned models indicate the shortcoming of treating all adjoining nodes for every relationship with equivalent significance. To beat this restriction, KB-GAT \cite{nathani2019learning} uses the concept of attention to recognize significant data in the locality of nodes. Like RGCN and SACN, the encoder-decoder approach is followed. It uses graph neural nets as encoders and ConvKB as decoders. In contrast to the prior approaches, it exploits GNNs to learn both relation and entity embeddings.

\subsection{Loss function layer}
Neural networks are specifically used to embed information in KGs. The inaccuracy of a predicate is determined using the explicit scoring potential for the embedding. The loss function is used in conjunction with the scoring function. Extensive research is underway to develop new scoring functions, yet until recently, there has been almost no emphasis put on researching novel loss functions \cite{mohamed2019loss}. The scoring elements of HolE and ComplEx are displayed similarly, although their effectiveness is opposite \cite{mohamed2019loss}. One possible explanation for this difference is the miscellaneous loss function used for the scoring energy function [ ]. Five distinct loss functions are mentioned in this study. Here $\lambda$ represents hyperparameter and $[t]^+$ mentions $max(t, 0)$. For Pointwise functions, given triplet $t$, if $t$ is true, $g(t)$ equals 1; otherwise, 0. For pairwise losses, $f(t')$ represents false fact and $f(t)$ equals true fact. 

Pointwise square error loss is used by RESCAL \cite{nickel2011three} where the objective is to limit the squared difference between the model scores and the labels. The ideal score for valid and false triples is 1 and 0, respectively. This loss profits by not having the hyperparameter contracting the space of hyper boundaries that differentiate with other loss functions. 
\begin{equation}
Pointwise \, square\, loss = \frac{1}{2} \sum_{t \in T}(f(t)-g(t))^2
\end{equation}

Pointwise hinge loss is used by HolE \cite{nickel2016holographic}. The objective is to limit the scores of negative realities and expand the scores of positive realities to a particular configurable value. In particular, this loss function reduces negative scores to - $\lambda$ and amplifies positive scores to + $\lambda$.
\begin{equation}
Pointwise \, hinge\, loss =  \sum_{t \in T}(\lambda-g(t)f(t))^+
\end{equation}

Pointwise Logistic Loss is utilized by CompleX \cite{trouillon2016complex} with the benefit of ignoring the parameter $\lambda$ results in a smoother loss slope. It likewise uses a logistic function to limit the negative triples score and advance the positive triples score. 
\begin{equation}
Pointwise \, logistic\, loss = \sum_{t \in T}log(1+exp(-g(t)f(t)))
\end{equation}

TransE \cite{bordes2013translating} and DistMult \cite{yang2014embedding} adopts the Pairwise Hinge Loss. Hinge loss can be executed in both pairwise or pointwise strategies. Linear learning is to rank to loss in order to maximize the difference or margin between true and false triplets.
\begin{equation}
Pairwise \, hinge\, loss = \sum_{t \in T^+} \sum_{t' \in T^-}(\lambda + f(t')-f(t))^+
\end{equation}

Lately, KGE methods have been developed to resolve the ranking issue as a form of multi-class characterization. A binary cross-entropy loss is employed by ConvE \cite{dettmers2018convolutional} to model the multi-class losses. It works by preparing the entire lexicon of entities for training each positive fact to such an extent that for a triple (s, p, o), all triples (s, p, o') with $o'$ belong to the entire set of entities with $o' = o$ considered false. Notwithstanding the extra computational expense of this methodology, it permitted ConvE to sum up over a bigger example of negative occurrences and outflank different methodologies. There are new loss functions proposed, for example, self adversarial \cite{sun2019rotate}, soft margin loss \cite{nathani2019learning},  multiclass negative log-likelihood \cite{lacroix2018canonical} ordinarily utilized for training KGE and have shown good performances in downstream tasks. It is important to focus on training different loss functions, just like scoring functions, when researching KGE for KGC. 


\subsection{Evaluation Metrics}
Typically, KGC task assessment metrics include Mean Reciprocal Rank, Mean Rank, and Hits@k \cite{akrami2020realistic}. They comprehensively assess the efficiency of KGC algorithms from various angles and are not complicated to use. 

Here, N refers to all the expectations.

Mean Rank works by calculating the average of rank associated with predictions between all competitors. The low value of the Mean Rank suggests the predictive power of the model is higher. The Mean Rank values can mirror the positioning of the right triples in the likelihood of setting up the test triples. In simple terms, it is a proportion of the exactness of the KGC algorithm.
\begin{equation}
Mean\, Rank = \frac{1}{|N|}\sum^{|N|}_{i=1} rank_{triple(i)}
\end{equation}

Mean Reciprocal Rank (MRR) works by predicting the scores of triples based on whether they are correct or not. It is an ordinarily utilized metric to quantify the impact of search algorithms. The larger the value of MRR, the better the performance of the model. On the off chance that the first anticipated triple is valid, its score is 1, and the subsequent true scores are 1/2, ---,1/n, where 'n' is the nth triple. The final score is the addition of all the scores. 
\begin{equation}
Mean\, Reciprocal\, Rank = \frac{1}{|N|}\sum^{|N|}_{i=1} \frac{1}{rank_{triple(i)}}
\end{equation}

Hits@K demonstrates the likelihood of the right prediction in the top K possible triples determined by the model. Hits@K addresses the capacity of the model to anticipate the connection between facts effectively.  In simple terms, it counts how many positive triples are ranked in the top k positions against a bunch of synthetic negatives. The value of K is generally chosen as 10. The value of Hits@k lies in the range of 0 and 1. The larger the value of the Hits score, the better the performance of the model.
\begin{equation}
Hits@K = \frac{1}{|N|}\, 1\, if rank_{triple(i)} =<K
\end{equation}



\section{Reinforcement Learning for Knowledge Graph Completion}

When employing KG to advance a question answering (QA) framework, only one of the triple is inferred to respond to the inquiry. For complex QA frameworks and when the given KG is mainly fragmented, it is essential to have the option to deduce obscure answers with existing triples. State-of-the-art embedding-based techniques limit their applications to model complex queries due to their inability to model the symbolic composition of facts present in KGs. An elective answer is to deduce missing links by orchestrating data from multi-hop paths, for example, BornIn(JoeBiden, Pennsylvania) $\wedge$ LocatedIn(Pennsylvania, USA) $\Rightarrow$ bornIn(JoeBiden, USA), given in figure 17. Reinforcement learning can help to understand questions and answers by modeling them as a sequential decision problem. As of late, the Path-Ranking Algorithm (PRA) \cite{lao2010efficient} arises as a promising technique for learning complex paths in huge KGs. PRA uses random walk with restart-based deduction component to run a depth first search to extract required features called relational paths. Although it works in a completely discrete space, making it hard to assess and look at comparable entities and relations in a KG.  

\begin{figure}[htbp]
\centerline{\includegraphics[width=7cm, height=3.5cm]{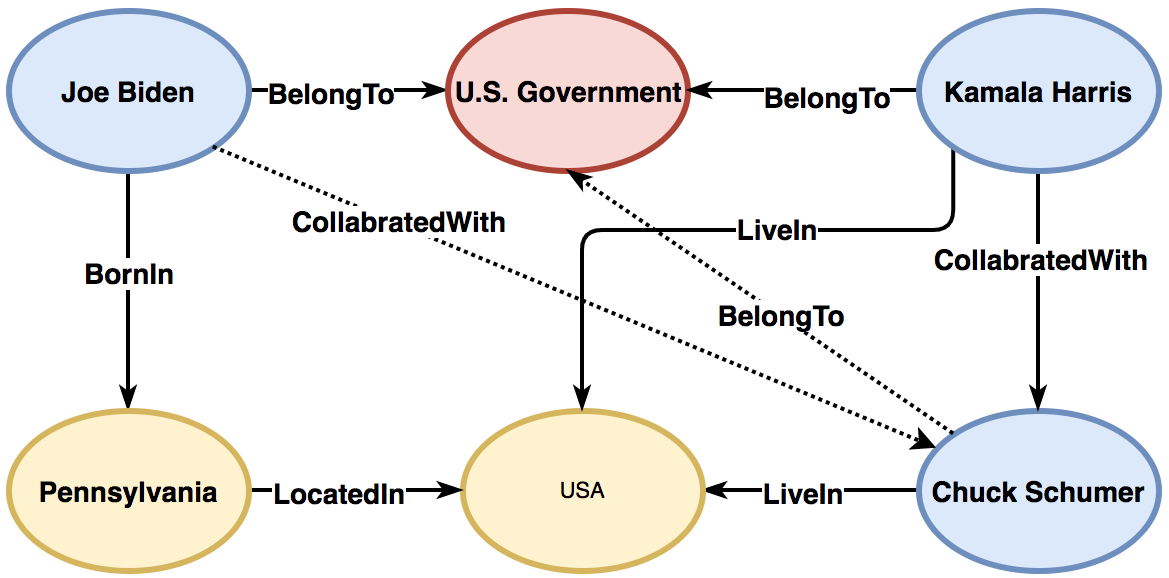}}
\caption{An example of knowledge graph completion to infer complex queries.}
\end{figure}

DeepPath \cite{xiong2017deeppath} is one of the earliest attempts to use reinforcement learning techniques to estimate multi-hop logic over a KG. The RL environment in DeepPath is defined by Markov Decision Process (MDP). The translation-based embeddings, namely TransE \cite{bordes2013translating}, and TransH \cite{wang2014knowledge} is utilized for encoding the state of the RL agent in low dimensional space, which is responsible for sampling the relations to extend the number of paths. The reward function for policy gradient-based training is defined to control the path search for better precision, accuracy, and productivity to better guide the agent. However, it needs to know the objective target entity ahead of time to direct the agent. DeepPath outperformed standard PRA and KGE methods, yet one principle issue is that its action space is moderately huge. Moreover, it must not be exploitative of complex tasks where the subsequent element is obscure and should be procured by inferring. Exploring the limitations of DeepPath, Das et al. \cite{das2017go} proposed MINERVA, which takes into account that the objective element found from the source element through the path. It does that without knowing the objective element and pre-figuring the path that worked on the past technique. It presents the Long Short Term Memory (LSTM) network to oversee previous paths and helps to guide the agent over a specific entity and relations that help better perform the policy-guided walk to the right element. 

In contrast to MINERVA, M-WALK \cite{shen2018m} utilizes RNN to map historical paths to the agent and Q values which then at that point employs the Monte Carlo sampling technique to unravel the direction. It lastly utilizes the Q-Learning algorithm to learn the values present in action. However, it exhibits two major drawbacks. RNN, as we know, suffers from a gradient explosion problem when the size of input data is significant. The Monte Carlo sampling typically requires a finalized direction to assess the technique and update the model, so the productivity is likewise low.

The walk-based QA frameworks may experience significant downsides. First, because most of the KGs have missing links, the agent may show up at a right answer whose connection to the source substance is absent from the KG without getting any reward. Second, since no ground truth is accessible for training, the agent might unintentionally navigate misleading paths that lead to the right answer. To counter this, Lin et al. \cite{lin2019multi} suggests two techniques for RL-based KG reasoning, namely, reward shaping and action dropout. The former one joins ability in demonstrating the semantics of triples with the representative thinking capacity of the path-based methodology. Whereas the latter one is more powerful in empowering the policy to test different paths. In most of the works at the decision level, only the absolute entity is named, by which the agent, in turn, brings about sparse and deferred rewards. To take care of this issue, Qiu et al. \cite{qiu2020stepwise} proposed a potential-based reward forming technique to supply extra compensations to the agent to direct its reasoning cycle, which can speed up the convergence and cause the model to perform better.  

Several works have been done to abolish the pre-training cycle needed to diminish the complexity. Specifically, the attention mechanism focuses harder on the neighbors to avoid choosing an invalid path. Wang et al. \cite{li2021memorypath} develop a system that joins LSTM and graph attention mechanism as memory parts to dispose of the fine-tuning process. Augmenting the work of \cite{lin2019multi} they further develop the deep reinforcement learning framework by introducing three techniques, namely, reward shaping, action dropout, and force forward. Recently, a methodology called DAPath \cite{tiwari2021dapath} is proposed dependent on distance-based reward in the RL system to map variable awards for various positions. To allow the model to recall the path that considers overseeing the memory of relations in the path, it considers GSA (graph self-attention) with gated recurrent unit (GRU). However, it experiences slow convergence and low learning proficiency. Most of the earlier methodologies are based on a popular REINFORCE \cite{williams1992simple} (policy gradient) algorithm, which as a rule has a huge variation and vigorously relies upon starting policy. 

Wang et al. \cite{wan2020reasoning} propose ADRL, which introduces the Actor Critic Algorithm that exploits policy gradient and sequence-based differential learning. Contrasted with REINFORCE, it works by updating the parameters in a single step fashion by utilizing the value function responsible for decreasing the variance of policy gradient without remaining idle so that the value function and policy gradient are trained concurrently. Less thought has been given to examining the hierarchical construction for KG reasoning, which exhibits performance improvement for modeling multiple semantics. Wan et al. \cite{wang2020adrl} introduce a framework called HRL (hierarchical reinforcement learning) that functions by augmenting the whole action into sub-actions. The component is carried out by a progression of the high to low-level policy. The former helps manage to learn historical data. The latter is answerable for learning sub-actions just as augmenting every action space into a light action space. Thus, the numerous semantics of each relationship can likewise be learned. 

\section{Comparative Analysis}
Looking at the execution of best-in-class models is precarious because of various training techniques adopted by researchers. For example, different loss functions (self adversarial or absolute margin), 1vsAll scoring or negative sampling, new types of regularization (for example, weighted and unweighted L1, L2, L3), initialization techniques, utilization of reciprocal relations \cite{kazemi2018simple}, and ablation studies are also not carried out.  Exploiting different techniques for KGE preparation makes it hard to analyze execution performance for different model designs, particularly when predictions are repeated in earlier examinations that utilized alternate training methods. For example, the parameters of a model are normally tuned utilizing grid search on a small search space by involving custom-built boundary parameters. A search space reasonable for one model might be imperfect for another. Of course, the more up-to-date training procedures can extensively execute better in performance \cite{salehi2018probabilistic} \cite{kadlec2017knowledge}.

In \cite{ruffinelli2019you}, summed up the effect of various model designs and distinctive training systems on model execution experimentally. They played out on a broad arrangement of trials utilizing mainstream model structures and methodology in a typical exploratory arrangement. As opposed to most earlier work, they considered training the models on enormous search space and performed model tuning utilizing quasi random search rather than grid search followed by bayesian parameter optimization. They tracked down the relative execution contrasts between different model structures and suggested that the results frequently dropped and sometimes improved compared with earlier outcomes. For instance, RESCAL \cite{nickel2011three}, which establishes the principal embedding model however is hardly explored to be in current works, showed solid performance and outflanked state of the art models, for example, TuckER \cite{balavzevic2019tucker}, and ConvE \cite{dettmers2018convolutional}. It recommends that appropriate training systems and hyperparameter settings shift essentially across data and models, demonstrating that a little change in search space can influence predictions on model execution.

Pouya Pezeshkpour et al. \cite{pezeshkpour2020revisiting} reconsidered and researched the current issues with evaluation metrics and clarified that the currently adopted techniques do not assess KGC, are hard to use for calibration, and cannot reliably differentiate between various models. Calibration is additionally a vital part of KGC that has as of late got consideration \cite{tabacof2019probability}. Safavi et al. \cite{safavi2020improving} show that calibration procedures can altogether diminish the alignment error of KGE models in the downstream tasks. Instinctively, calibration is a post-preparing step that changes KGE expectation scores to illustrate real and correct probabilities. \textit{“Treating the likelihood of truth of a triple $(\sigma(\psi(s, r, o))$ for triple s, r, o) as the certainty of the model for the triple, the model is considered to be calibrated if the certainty lines up with the pace of confirmed realities”} \cite{pezeshkpour2020revisiting}. If certainty is equivalent to 0.5, then around half of triples with this certainty to be valid. Assuming this extent is a long way from 50\%, the model isn't calibrated, i.e., the model is underconfident if the extent is greater and overconfident if it is lower. Under an optimal circumstance, if the model predicts that a triple is valid with a 0.9 certainty score, it ought to be right 90\% of the time. Model calibration needs dependable certainty estimation and successful alignment strategies to fix the calibration mistakes. As far as KGE is concerned, two certainty or confidence estimation techniques, SigmoidMax (SIG) and TopKSoftmax (TOP), are exploited in \cite{tabacof2019probability} \cite{safavi2020improving}. 

There are two calibration techniques. Isotonic regression \cite{zadrozny2002transforming} is a non-parametric technique and does take into account sigmoid assumption. It fits an increasing constant function to the model yield and is generally suitable for many examples. However, it is prone to overfit. Platt scaling \cite{platt1999probabilistic} on the other takes into account the sigmoid function that learns scalar weights to yield a confidence score for each example. It might work better for smaller datasets. Model Calibration has a few advantages. According to the framework's point of view, language processing pipelines that incorporate KG can depend on calibrated scores to determine which KGE forecasts to trust. According to a research point of view, incorporating calibration helps determine the output predictions for acknowledging KGE models. As given in figure 18, we can easily analyze how a very much calibrated model resembles. A straight spotted line addresses an ideal calibrated model though the red shaded line addresses an uncalibrated model. 

\begin{figure}[htbp]	
	\centering
	\begin{subfigure}[t]{1in}
		\centering
		\includegraphics[width=1in]{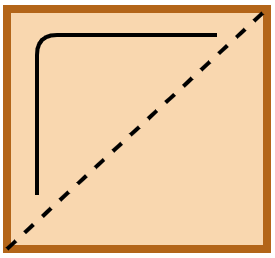}
		\caption{}\label{fig:1c}		
	\end{subfigure}
	\quad
	\begin{subfigure}[t]{1in}
		\centering
		\includegraphics[width=1in]{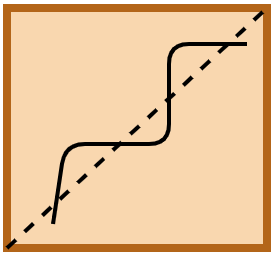}
		\caption{}\label{fig:1b}
	\end{subfigure}
	\caption{(a) Uncalibrated Model (b) Calibrated Model}\label{fig:1d}
\end{figure}

Models after calibration procedures work impressively better compared to uncalibrated strategies \cite{safavi2020evaluating}. 
Nonetheless, two primary issues are confining the viability of the probabilistic calibration techniques for link prediction tasks \cite{chami2020low}. Firstly, the lack of suitable confidence estimation. The evaluation techniques \cite{} revolve around the ideal score and contrast it with different scores in the score sequence. High ideal scores lead to high certainty yet do not accomplish a similar level of precision since the score of each triple shows its relative ordering among other triples in a single prediction. Secondly, the inconsistent and unreliable calibration metrics. Expected Calibration Error (ECE) \cite{niculescu2005predicting} is ordinarily used to assess the impact of calibration, yet is not appropriate for downstream tasks such as link prediction. Benefitted from a causal inference analysis, Kai Wang et al. \cite{wangneighborhood} proposed a novel neighborhood intervention consistency (NIC) method that can effectively intercede the scoring cycle of KGE models. In particular, it creates a progression of neighborhood vectors for an input element by changing the entity vector in various dimensions and inspecting whether the model's output changes or matches the initial one. On this premise, the authors also designed neighborhood intervention values and a dimension selection system for high-dimensional KGE models to focus on efficiency. However, there is a tradeoff between the number of dimensions incorporated and the accuracy achieved. Selecting a proper neighborhood is very much needed to focus on both efficiency and predictive power.

\section{Towards Embedding Real World Knowledge Graphs}

Existing methodologies fundamentally revolve around static link structure between a finite arrangement of entities overlooking the assortment of information types that are regularly utilized in information bases like content, arithmetic values, images, uncertain and temporal information. In this section, a higher outline on utilizing this real-world knowledge on link prediction tasks is discussed. As shown in figure 19, this segment is split into three subsections, multimodal knowledge graph, temporal knowledge graph, and uncertain knowledge graph embeddings. 

\begin{figure}[htbp]
\centerline{\includegraphics[width=8cm, height=5cm]{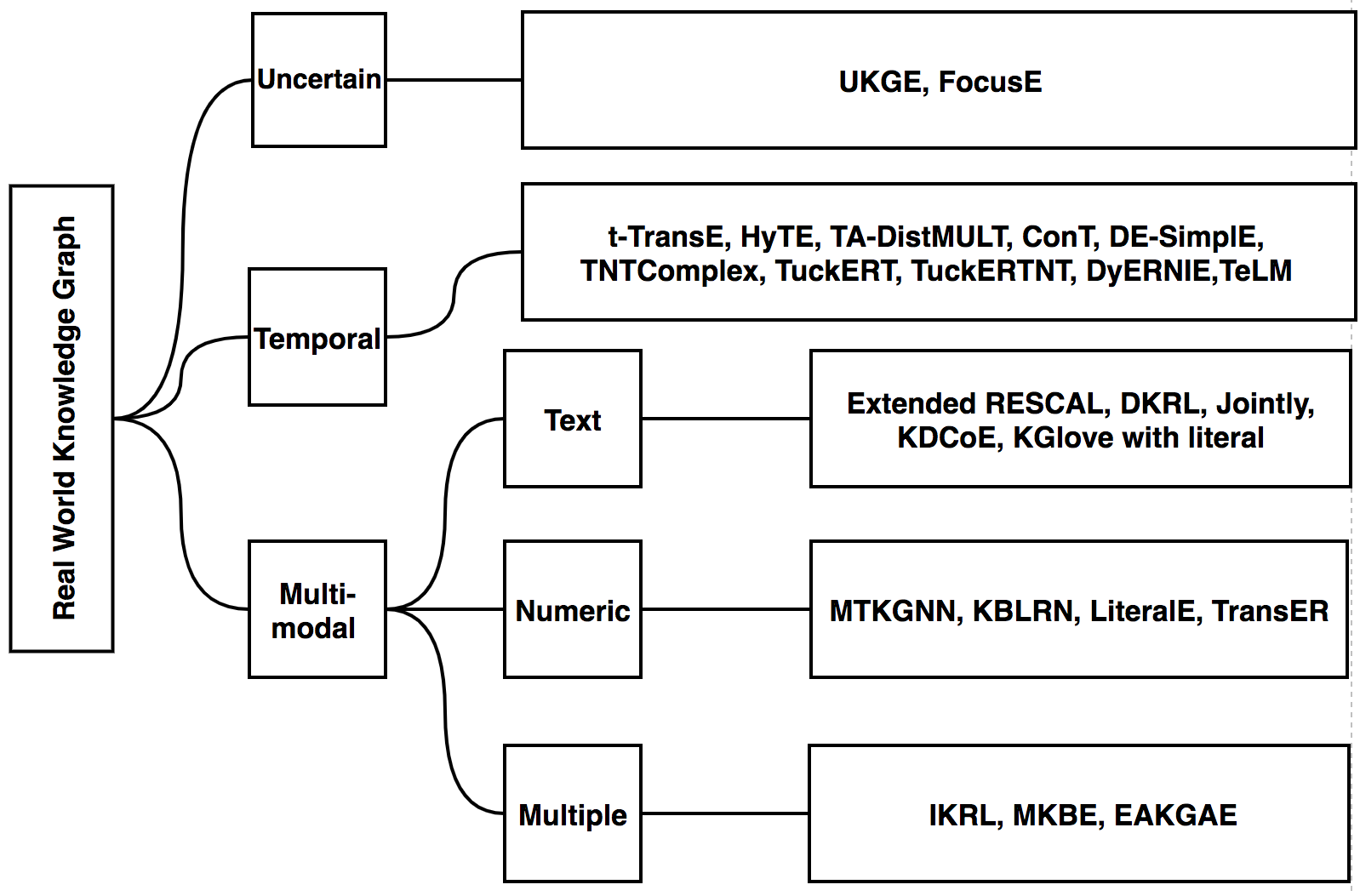}}
\caption{Hierarchical Distribution for emneddings real world knowledge graphs.}
\end{figure}

\subsection{Multimodal Knowledge Graph Embeddings}

Aside from relations to a fixed arrangement of triples, multimodal knowledge bases not solely incorporate mathematical attributes like age, monetary, and geo-data, but inherit literary features like names, biodata, and designation, and images such as profile photographs, banners, etc. These knowledge types act as a critical part providing additional information that can play as a catalyst for improving the accuracy in knowledge graph completion tasks. For instance, the literary descriptions and images may give proof of an individual's age and occupation. If we consider a multimodal KG containing an image of a person with a description as text, a ‘Person’ designation can easily be attributed utilizing a literal image, while the description contains his identity. Incorporating multimedia into existing methodologies as triples are difficult as they relegate every element to a particular vector and anticipate missing connections or characteristics by specifying the potential values, the two of which are only conceivable if the elements come from a small-scale enumerable set. In this study, the KGE models with multimodality are separated into the accompanying classes dependent on the multimedia used: Numeric, Text and Multiple literals. A KGE model which leverages no less than two categories of multimedia is considered multimodal.

\subsubsection{Text Literal}

Extended RESCAL, DKRL, Jointly,  KDCoE and KGlove with literals leverage the text literals. Extended RESCAL \cite{nickel2012factorizing} augments the initial RESCAL approach by handling textual attributes more effectively and managing the sparsity of the tensors. The idea was to utilize multiplicative update rules to extend the nonnegative factorization such that the triplets that contain textual knowledge are encoded in a matrix by tokenizing and stemming operations. This model, however, converges slowly compared to RESCAL and does not take into account the grouping and sequencing of the text. 

DKRL (Description Embodied Knowledge Representation Learning) \cite{xie2016representation} is an extension of TransE that constructs embeddings of triplets by joining structure and description features. Deep CNN and a continuous bag of words (CBOW) are utilized to produce the description-based representation of entities. The structural highlights are then incorporated by translation-based scoring function TransE. However, it isolates the objective functions into energy elements of description and structural representation, which is inefficient. Instead of utilizing CNN, the literature suggests Jointly \cite{xu2017knowledge} as it exploits attentive LSTM and incorporates the textual as well as structural knowledge of entities into a joint representation. However, the scoring function in both DKRL and jointly is based on TransE only.

KDCoE \cite{chen2018co} centers around the formation of an arrangement for elements of multi-lingual knowledge bases by making novel inter-lingual links with high confidence. It employs a multilingual KG for semisupervised cross-lingual training and performs cotraining of a multi-lingual KGE model with a multi-lingual description embedding model that repetitively associates with each model for entity descriptions and structured based knowledge. Michael Cochez et al. \cite{cochez2018first} proposed KGlove with literals intending to incorporate descriptions on their original KGlove \cite{cochez2017global} model. The two co-occurrence matrices are generated autonomously and are finally merged to perform a joint embedding. The first co-occurrence matrix is based on the KGlove approach, which performs the individualized page rank on the original weighted graph and is normalized using the optimization used in Glove. The latter is an inverted edged graph, and the named entity recognition is performed before the former matrix is constructed and subsequently normalized.

The fundamental difference between these models lies in the procedures used to get as much benefit from the information in the text as possible. The advantage of KDCoE over other models is that it considers the descriptions present in multi-lingual graphs. However, the different types of text literals are not widely used in the aforementioned literature because the works are more aligned towards longer text literals and less overview on shorter text like labels and names.

\subsubsection{Numeric Literal}

The works that leverage numerical knowledge are MTKGNN, KBLRN, LiteralE, and TransEA. MTKGNN \cite{tay2017multi} adopts both relational networks and attribute networks to train triple classification and regression tasks, respectively, to obtain the knowledge contained in entities and learned embeddings. A linked fact is fed into a non-linear transformation in a relational network and thereafter implements a sigmoid function to obtain linear transformation. Two regressions are performed for the head and tail properties separately to predict continuous attribute values in attribute networks. Finally, both networks are modeled in a multifunctional design using a common embedding space. Mathias Niepert et al. \cite{garcia2017kblrn} propose a novel KBLRN approach to merge relational, latent, and numerical highlights to represent large numerical values. The Probability of Experts (PoE) method is utilized to merge these features and train them together from start to finish. 

LiteralE \cite{kristiadi2019incorporating} feeds data into the current latent feature model by adjusting the scoring function in base model DistMult. It does by supplanting the vector representation of the elements in the scoring function with the literally enriched entities.
To produce new entities vectors that are lexically rich, it exploits a learnable transformation method that inputs the original entities and its initially aligned literal vectors as data sources and maps them to new vectors.  TransEA \cite{wu2018knowledge} on the other hand, is comprised of two embedding models, namely TransE and Attribute Embedding Model (AEM). AEM undergoes linear regression with attributive numeric features as input. The TransE is then characterized using the individual loss measure of the part model with thresholds for the weights to be distributed.

Despite their commitment to leverage numerical literals, each method discussed neglect to understand the meaningful relationship between the information and a variety of multimodality. For example, 'Year 2010' and '2010e' can be seen as something similar because the type semantics are discarded. Also, the standardization is not implemented properly. Henceforth the semantic equivalence between two values, for example, '500nm' and '5m', is not highlighted. Similarly, most models do not have the appropriate components to deal with multi-valued literals.

\subsubsection{Multiple Literals}

To the best of our knowledge, MKBE \cite{pezeshkpour2018embedding} is the current state of the art model that incorporates numeric, text and image multimedia for modeling KGE. It works on the principle of DistMult which adds neural encoders to different types of information to form embeddings for triplets. A fixed-length vector is encoded utilizing CNN for image knowledge type. For text features, LSTM is employed to learn and extract the sequences present in text data. The scoring used in MKBE is the same as the scoring function employed in DistMult and is utilized to decide the accuracy of likelihood of triples. IKRL \cite{xie2016image} incorporates TransE for structural learning and mutual training with an image encoder to create embeddings in each instance for the image relation. It additionally configures the attention of each instance of the image using multi-valued attention based training. Other approaches like EAKGAE \cite{trisedya2019entity} and \cite{xie2016representation} are also worth mentioning. However, they only include text and numeric literals.

\subsection{Temporal Knowledge Graph Embeddings}

A few KGs include temporal realities, for example, the triple \textit{(DonaldTrump, PresidentOf, USA)} only substantial in a particular time span (2017, 2021). Temporal Knowledge graphs like YAGO3 \cite{suchanek2008yago}, ICEWS2014, ICEWS2005-15, ICEWS \cite{boschee2015icews}, GDELT \cite{leetaru2013gdelt} feed time data into facts. Triples appended with time data are addressed as quadruples, formed like \textit{(s, r, o, T)}, where T signifies the timestamp. Conventional KGE models dismiss time data, prompting an insufficiency of performing link prediction on Temporal KGs, including relations, for example, \textit{(?, PresidentOf, USA, 2018)}. 

\begin{figure}[htbp]
\centerline{\includegraphics[width=7cm, height=5cm]{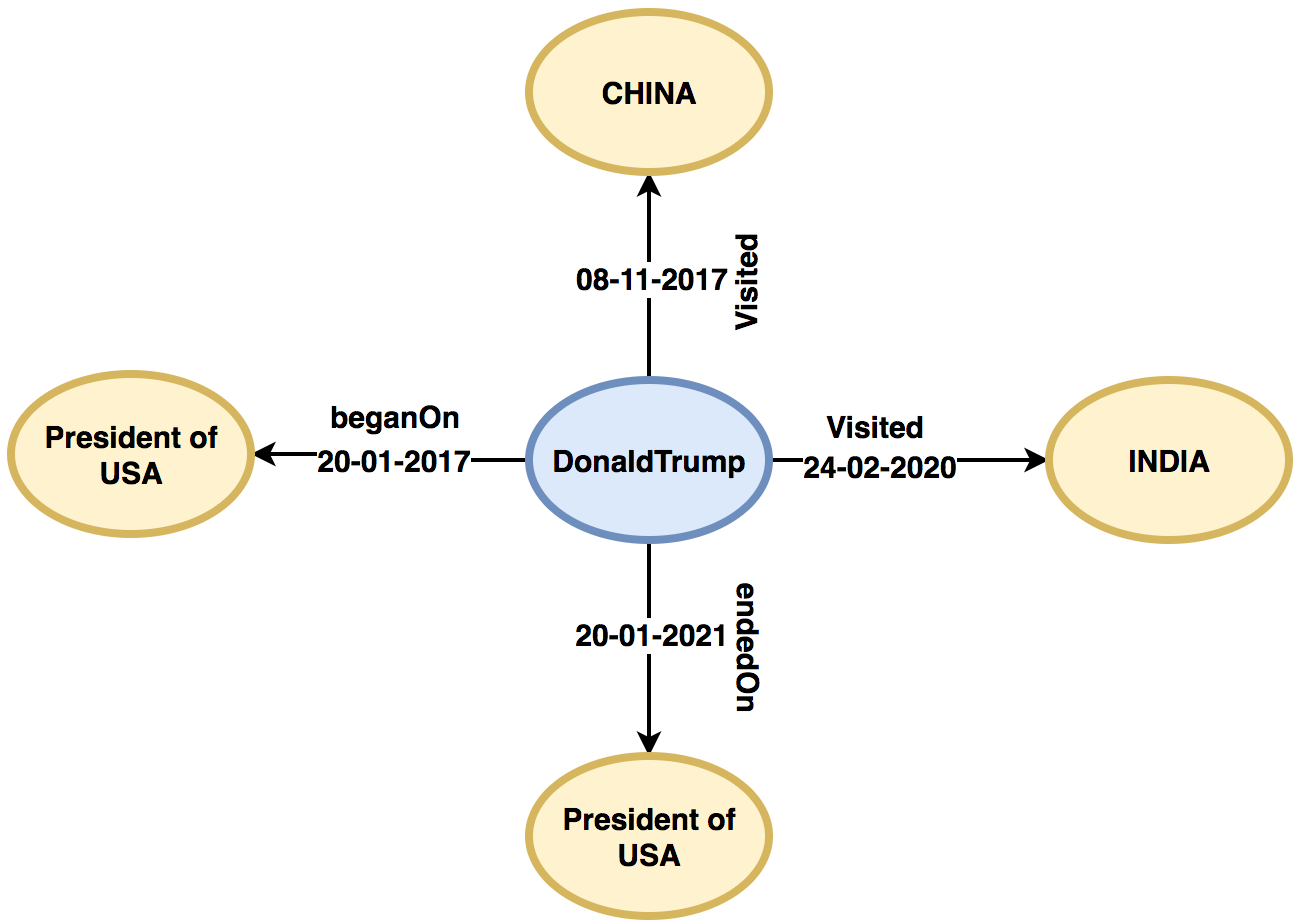}}
\caption{An example of Temporal Knowledge Graph.}
\end{figure}

In general, KGE models neglect to undertake time-based modality while learning embeddings for static Knowledge bases. Temporal-KGs is a significant yet infrequently examined research area. The past  few years have witnessed a good number of TKGs models \cite{jiang2016encoding} \cite{dasgupta2018hyte} \cite{garcia2018learning} \cite{lacroix2020tensor} \cite{han2020dyernie}. The purpose of the temporal KGE is to dynamically change the (time-based) associated relationships between adjacent triples. t-TransE \cite{jiang2016encoding} is a combined model of TransE and temporal boundaries. It suggests placing these bounds in temporal order on the vector space to maintain temporally stable and precise embeddings. Influenced by TransH, HyTE \cite{dasgupta2018hyte} associates entities with temporal knowledge and relation by projecting them into a hyperplane trained by temporal data and achieving embeddings by limiting translation distances. To exploit sequences present in temporal knowledge, TA-DistMult \cite{garcia2018learning} adopts RNN to gain efficiency with the time-based features of the relation, which is used in TransE and DistMult for link prediction tasks. 

To leverage the cognitive capacity, ConT \cite{ma2019embedding} generalizes static knowledge graph embedding methods, including RESCAL and Tucker, to episodic tensors (Et) in order to reduce complexity. Tree and ConT are two novel speculations of RESCAL to Et. ConT acquires exceptional execution for modeling temporal Knowledge graphs due to its latent representational flexibility of sparse tensor for time index. Even though the complexity of Tree and ConT is decreased when contrasted with episodic Tucker, it may, however, cause fast overfitting during training. DE-SimplE \cite{goel2020diachronic} incorporates diachronic word embeddings responsible for deriving entities' characteristics by uncovering the implications of advanced information over time. It simply works by combining it with a static KG such as SimplE for the function of temporal KGC.

Inspired by the CP (canonical product) decomposition of the order 4 tensor, TNTComplEx \cite{lacroix2020tensor} presents an extension of ComplEx for Temporal KGC. Although TNTComplEx achieves significant functioning, it is difficult to accurately determine the position and rank of tensors using CP decomposition. Propelled by the TuckeR decomposition of order 4 tensor, the literature proposed TuckERT \cite{shao2020tucker}, a variation of TNTComplEx. However, this model does not manually select the position of the tensor, and it considers only temporal facts. TuckERTNT \cite{shao2020tucker}, an extension of TuckERT \cite{shao2020tucker}, is additionally proposed in literature to include both static and temporal facts.

Temporal KGs often exhibit various concurrent non-Euclidean features, such as hierarchical and cyclic designs. The current KGE approach to the representation of temporal knowledge of entities and their dynamic progression in Euclidean space cannot fully capture such specific structures. To curb this, DyERNIE \cite{han2020dyernie}, a non-Euclidean embedding approach has been proposed that masters developing entities as a result of Riemannian manifolds, where structures are assessed from the geometrical bends of hidden information.
But this approach is highly dependent on a distance function, which may hamper the performance. More recently, a new temporal KG completion method TeLM \cite{xu2021temporal} is proposed. It performs order 4 tensor factorization of temporal knowledge using a linear temporal regularization for modeling time-based embeddings. It also uses multi-vector embeddings to represent knowledge as it provides better generality and greater expressivity for temporal KGEs as opposed to single and complex-valued embeddings.

\subsection{Uncertain Knowledge Graph Embeddings}

Knowledge Graphs, for example, ConceptNet \cite{speer2017conceptnet}, carry uncertain data with a certainty score allocated to each triplet. In a numeric-enhanced KG, each triple is associated with a numeric feature. It is significant to add a triplet with a particular numeric feature semantics, as these numbers may encode significance, vulnerability, strength, and so forth.  For instance, figure 21 suggests that numeric highlights demonstrate the significance of a relationship. The triple (ANDREW, Skill, MLOps, 0.90) is accordingly more significant than (ANDREW, Skill, Physics, 0.15) as far choosing a career path is concerned.

\begin{figure}[htbp]
\centerline{\includegraphics[width=8cm, height=5cm]{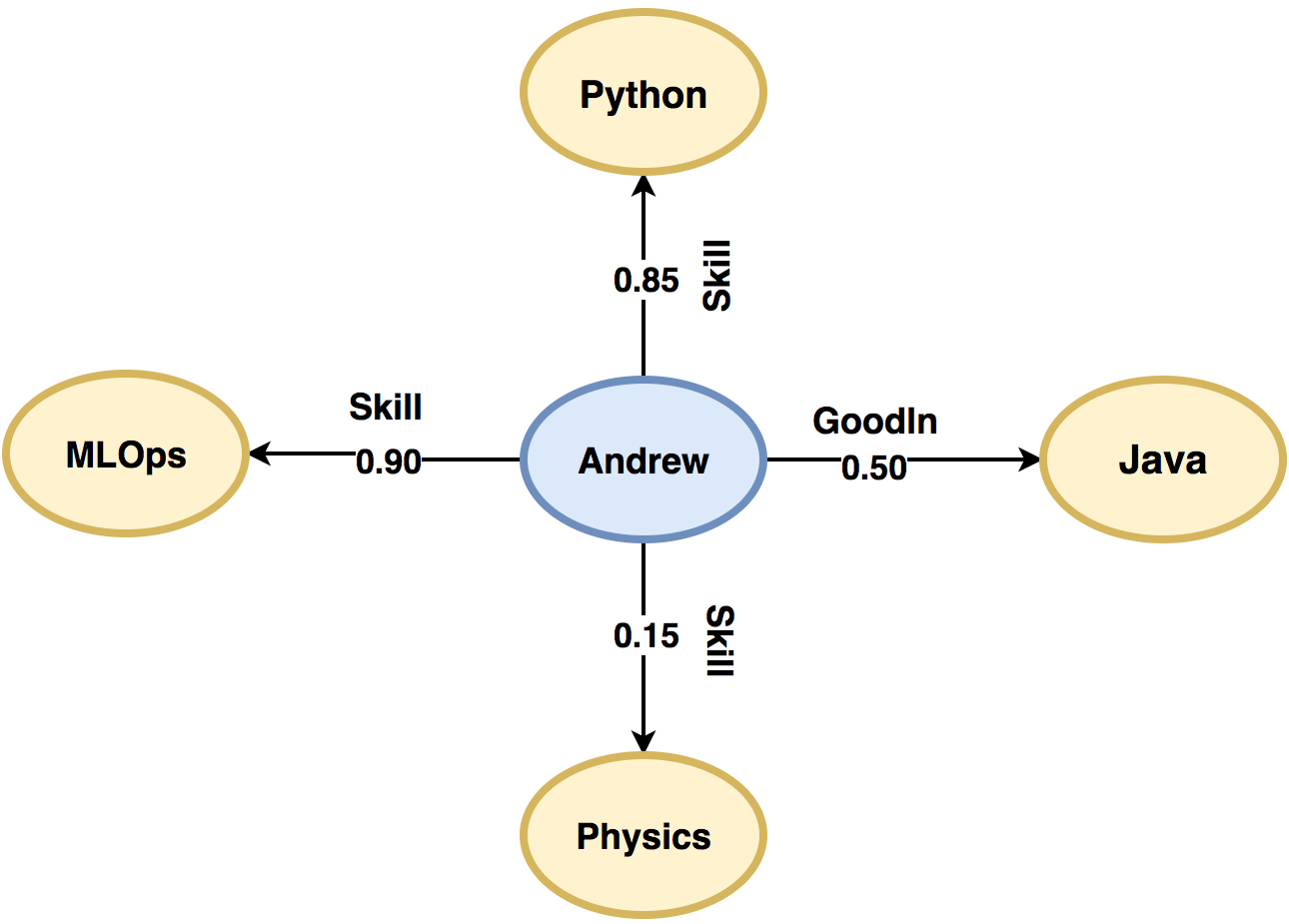}}
\caption{An example of Uncertain Knowledge Graph}
\end{figure}

Many works are mentioned in the literature of knowledge graph representation learning that supports multimodal data and leverage numeric features related to node entities to produce better embeddings for upgraded link prediction tasks \cite{kristiadi2019incorporating} \cite{wu2018knowledge} \cite{pezeshkpour2018embedding}. However, these models are not intended to gain from numeric qualities related to edges in a KG. With the prominent exemption of \cite{chen2019embedding}, supporting numeric features is up to date still an under-achieved exploration. UKGE \cite{chen2019embedding} produces certainty scores for seen triples by flattening the numeric values from range 0 to 1. It then utilizes probabilistic logic \cite{kimmig2012short} to anticipate the likelihood estimates for triples that are unseen by mutually preparing a model to regress over the certainty scores. A constraint of this methodology is that out-of-band logic principles are needed as extra information. It is additionally significant to note that UKGE targets at supporting uncertain KG, in particular graphs whose edge numeric features address uncertainty.  Sumit Pai et al. \cite{pai2021learning} proposed FocusE, an extra layer that fits between the loss and scoring layers of an ordinary KGE technique, and it is intended to be utilized during training. In contrast to conventional techniques, prior to inputting the scoring layer to a loss function, they balance the output dependent on numeric attributes to acquire focused scores. They influence numeric values so that while preparing the model, it focuses more on the triples whose associated numeric values are large. However, this extra layer does not exploit unseen triples into consideration, and incorporating multi-valued numeric features is still a challenge.

\section{Software Libraries}

In this section, we will briefly discuss the software ecosystem that revolves around embedding knowledge graphs. Over the past three years, several libraries have been distributed recently with plans to upgrade further the development process related to KGE (as shown in figure 22). The research and development community has additionally distributed code to work with open source contributions. A correlation between open-source software libraries based on features, accessibility, scalability, state of the art produced, and programming environment practices is reviewed in this paper.

\begin{figure}[htbp]
\centerline{\includegraphics[width=7cm, height=2.5cm]{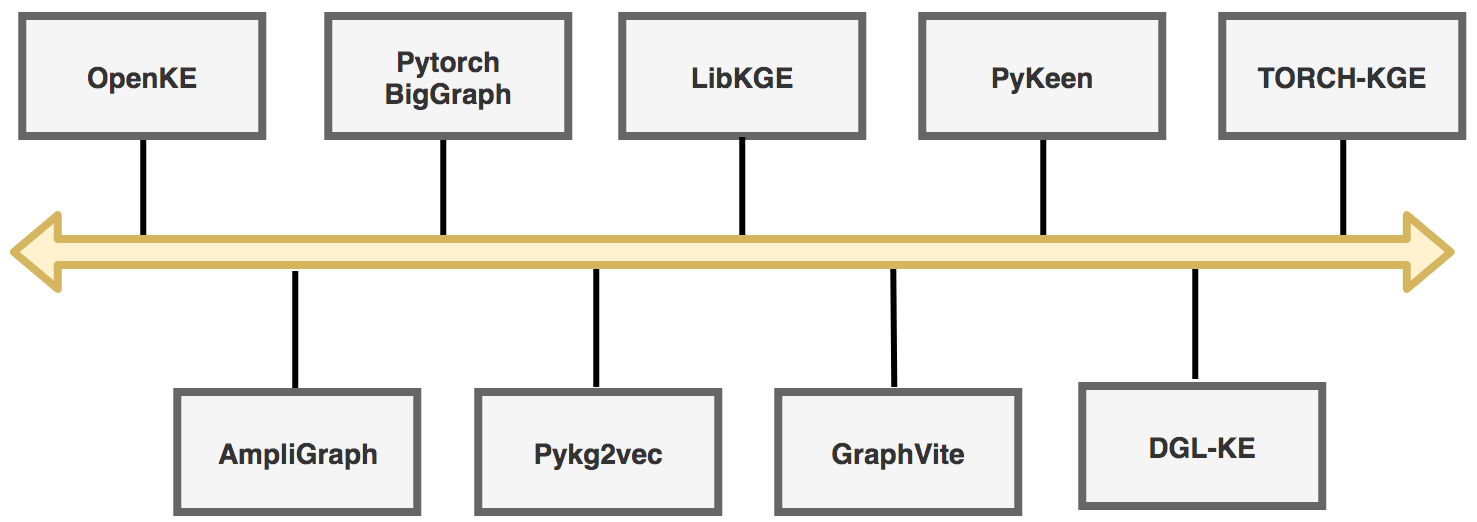}}
\caption{Open source libraries for training KGE.}
\end{figure}

The center of the KGE's are models, and we have mentioned (in section 5) that there are many models available for training the KGE. The accessibility of models for each library is recorded in Table 2. Some of the most common models are TransE \cite{bordes2013translating}, CompleX \cite{trouillon2016complex} and DistMult \cite{yang2014embedding} accessible practically in almost all the libraries and there are a few varieties present across the rest of the models for which library gives what models. Libraries like Pykg2vec \cite{yu2021pykg2vec} and PyKEEN \cite{ali2021pykeen} offer a wide scope of models, for example, RGCN \cite{schlichtkrull2018modeling}, TuckER \cite{balavzevic2019tucker}, NTN \cite{lin2015learning}, KG2E \cite{he2015learning} which are absent in libraries such as AmpliGraph, Pytorch BigGraph and GraphVite etc. It all depends on what the use case is, if it is a research task where different models or approaches need to be analyzed then the libraries with the largest number of models are ideal. However, assuming that the requirement is good performance only for applicable use cases, it may not be fundamentally relevant to have too many models where we only need high performance and accuracy.

\begin{table*}[h]
\scriptsize
\caption{A comparative analysis of open source libraries for training KGE. S1*: TransE, DistMult, ComplEx, TransH, TransD, TransR, RESCAL, HolE, SimplE. S2*: KG2E, NTN, ProjE, TuckER. S3*: TransE, DistMult, ComplEx}
\begin{center}
\begin{tabular}{|P{2.2cm}|P{3.5cm}|P{2.0cm}|P{2cm}|P{2cm}|}
\hline
\textbf{Library}& \textbf{Model}& \textbf{Pre Training support}& \textbf{Performance}
& \textbf{Framework}\\
\hline
 OpenKE\cite{han2018openke}&S1*, Analogy& WikiData, FreeBase& Support GPU& Pytorch, Tensorflow \\
\hline
 AmpliGraph \cite{costabello2019ampligraph} & S3*, HolE, ConvKB, ConvE & - & Support GPU& Tensorflow \\
\hline
 Pytorch BigGraph \cite{lerer2019pytorch}&S3*, RESCAL& WikiData& Support GPU, CPU (Distributed Execution)& Pytorch\\
\hline
 Pykg2vec\cite{yu2021pykg2vec}&S1*, S2*, Analogy, CP, ConvE, RotatE, TransM, SLM, SME, ComplexN3 & - & Support GPU& Pytorch, Tensorflow  \\
\hline
 LibKGE \cite{broscheit2020libkge}&S3*, RESCAL, SimplE, ConvE, RotatE, TuckER, CP   & WikiData,  FreeBase& Support GPU& Pytorch \\
\hline
 GraphVite \cite{zhu2019graphvite}&S3*, SimplE, RotatE, QuatE& WikiData& Support GPU, CPU (Distributed Execution)& Pytorch \\
\hline
 PyKeen \cite{ali2021pykeen}&S1*, S2*, ConvKB, ConvE, RotatE, RGCN, SME& - & Support GPU& Pytorch \\
\hline
 DGL-KE \cite{zheng2020dgl}&S3*, TransR, RESCAL, RotatE& - & Support GPU, CPU (Distributed Execution)& Pytorch\\
\hline
\end{tabular}
\end{center}
\end{table*}

Access to pre-trained embeddings for large-scale knowledge graphs to perform downstream tasks is an important perspective. This is probably the most useful parameter when the client does not have enough resources and technology to train KGE's from scratch. Thus, utilizing Pre-Trained models that are accessible in certain libraries for instance, WikiData in (OpenKE \cite{han2018openke}, pytorch BigGraph \cite{lerer2019pytorch}, GraphVite \cite{zhu2019graphvite}, Lib-KGE \cite{broscheit2020libkge}) and a few parts of FreeBase is likewise accessible in OpenKE. Libraries, for example, AmpliGraph \cite{costabello2019ampligraph} and Lib-KGE \cite{broscheit2020libkge} additionally offer benchmark datasets.

Sometimes, the user may need to train embeddings themselves, so it is important to consider scalability when we cannot leverage pre-trained embeddings. If the graph is enormous, we may need to scale it through distributed execution offered by Pytorch BigGraph, GraphVite, and DGL-KE. Another important viewpoint is the core framework; that is, different libraries support different frameworks like TensorFlow and PyTorch. A large subset of them support PyTorch; However, if the user only knows TensorFlow, it may be more appropriate to use libraries that maintain TensorFlow, such as AmpliGraph and Pykg2vec. 

There are likewise other extra features of libraries that should be looked at. The most significant are detailed hereafter in the following subsections. 

\subsection{OpenKE}
OpenKE \cite{han2018openke} assembles a cohesive and coordinated framework for organizing information and memory. It likewise implements GPU and parallel learning to accelerate model training. It binds together numerical structures for certain models and encapsulates them to be maintained with modular programming. Some highlights are carried out in C++ to maintain extensibility. More KGE models are required to keep up with the stable embeddings of some huge scope KG. It is additionally tough to use as it has no versioning and is not appropriately contained for any package manager.

\subsection{AmpliGraph}
AmpliGraph \cite{costabello2019ampligraph} gives instinctive APIs that are intended to decrease the code sum needed to learn models that foresee triples in KG. It depends on TensorFlow and is intended to run flawlessly on CPU and GPU units to accelerate model training. It contains modules that help load KG and provides knowledge discovery API to find new triples, group elements, and explore duplicates. It is also well documented and provides slack support with google colab instructional exercises.

\subsection{PyTorch BigGraph}
PyTorch BigGraph \cite{lerer2019pytorch} was created to circumvent the issue of scalability in huge KG. It provides distributed training on a set of machines and accomplishes this through the segmentation of graphs. Therefore, the models do not need to be stacked entirely in memory. Multithread computation on each machine is working on different subsets of the graph simultaneously with distributed execution across different machines. This enables learning on huge graphs whose embeddings will not fit in a single GPU unit, yet it cannot provide top-notch embeddings for small graphs without careful tuning.

\subsection{Pykg2vec}
Pykg2vec \cite{yu2021pykg2vec} provides functions for hyperparameter optimization such as Bayesian optimization with negative sampling techniques in batches, as well as information charts to visualize embeddings and model performance metrics. The mini-batches generated for negative sampling result from employing multiple concurrent cycles that smoothens the batch generation process. These mini-batches are then transferred to a queue to be handled by the model carried out in TensorFlow. The batch generator function runs autonomously so that there is low idleness for faster execution. It likewise pictures the latent portrayals of triples on the 2D plane utilizing t-SNE \cite{van2008visualizing}, which helps to investigate the model for training KGE's. 

\subsection{LibKGE}
The decision of training methodology and hyperparameters are more powerful for model execution regularly more than the class of model itself. LibKGE \cite{broscheit2020libkge} intends to give clean executions for parameter search, assessment techniques, and training that can be utilized with any model. For tuning of parameters, it upholds Bayesian Optimization, grid search, and quasi-random search. Each preparation work or parameter search can be hindered and continued at any time. LibKGE benefits further from its configuration because everything can be treated as a hyperparameter, even the decision of the model and the score function. It performs broad logging in both human and machine-readable formats during analyses and screens execution metrics, such as runtime, memory use, loss, and evaluation metrics.

\subsection{GraphVite}
GraphVite \cite{zhu2019graphvite} extends existing embedding techniques for GPUs and essentially speeds up training to generate embeddings on a single unit. It is centered around multi-GPU and does not support distributed training. It creates a subgraph, moves all the information in the subgraph to GPU memory and performs several mini-batch steps on the subgraph. This strategy minimizes information development between the CPU and GPU at the cost of reducing the liveliness of the embedding, which generally results in much slower convergence.

\subsection{PyKEEN}
PyKEEN \cite{ali2021pykeen}, one of the early programming bundles for KGE modeling, had some issues such as the model should only be modeled under the stochastic neighborhood closed world method. The evaluation technique for large KGs was excessively delayed, and was intended to be used only via the command line interface line. PyKEEN (Python KnowlEdge EmbeddiNgs) 1.0 is a refreshed form that controls the previously mentioned issues and empowers clients to create KGE models dependent on a wide scope of association models, loss functions, distinctive training approaches, and grants the modeling of converse relations. A programmed memory streamlining step is executed that registers the configuration of models and hardware availability to assess batch sizes for the current model arrangement prior to executing the real experimentation. If the user input batch size is too large, then the programmed memory optimization decides the largest sub batch size for execution. Through the integration of Optuna, extensive HPO functionalities are also included.

\subsection{DGL-KE}
DGL-KE \cite{zheng2020dgl} introduces distinct new advancements that accelerate execution on KGs with large numbers of nodes and billions of edges using distributed parallelism, multiprocessing, and GPUs. These improvements are intended to build information around, reduce overhead, cover calculations with memory ingress, and achieve high efficiency. This bundle is implemented with Python on top of the Deep Graph Library (DGL) \cite{wang2019deep} with a C++ based operable key-value store explicitly intended for DGL-KE. Various negative inspection procedures have also been incorporated to create smaller than expected clusters with some embeddings engaged in the batch, which minimizes information movement from memory to processing units.

\subsection{TORCH-KGE}
TORCH-KGE is the most recent software package that has been acknowledged at KDD, which is worth mentioning. It \cite{boschin2020torchkge} is a Python module for KG that completely depends on PyTorch. Its major strength is a rapid evaluation module for link prediction. Extensive consideration is given to improved code proficiency and simplicity, documentation, and API stability.

\section{Open Research Challenges}

\begin{table*}[h]
\scriptsize
\caption{Towards Open Research Challenges for Knowledge Graph Completion}
\begin{center}
\begin{tabular}{|P{2.0cm}|P{1.7cm}|P{0.6cm}|P{8.0cm}|}
\hline
\textbf{Challenge}& \textbf{Model}& \textbf{Ref.}& \textbf{Description}\\
\hline
Interpretability& CRIAGE &  \cite{pezeshkpour2019investigating}& Defense mechanism for the automatic identification of a triplet from a graph that may alter the prediction for a target fact by adversarial modifications. \\
\cline{2-4}
& - &\cite{bhardwajpoisoning} & Scalable attack mechanism to effectively generate poisoning attacks for KGE models. \\
\cline{2-4}
& IDOpt, IDRank  &\cite{zhou2019adversarial} & Defense mechanism for similarity-based link prediction by effectively modeling target links as a Bayesian Stackelberg game.\\
\hline
Few Shot Learning& CogKR &\cite{du2019cognitive}& Cognitive graph that incorporates dual theory in cognitive science into a reasoning module for one-shot learning.\\
\cline{2-4}
& FSRL &\cite{zhang2020few}& Adopt heterogenous graphs for KGC by utilizing recurrent auto-encoder aggregation. \\
\cline{2-4}
& FTAL &\cite{mirtaheri2020one}& Incorporate the self-attention method to encode temporal dependence and a network to capture similarity scores for temporal graphs\\
\cline{2-4}
& GMatching &\cite{ xiong2018one}& One shot learning technique to learn entity embeddings utilizing local neighbor encoder.\\
\hline
Hyper Parameter Search&pykg2vec &\cite{yu2021pykg2vec}& Provides built-in automatic Bayesian hyperparameter optimization module to detect optimized hyperparameters for model training.\\
\cline{2-4}
& LibKGE &\cite{broscheit2020libkge}& Support quasi random search, grid search, manual search and bayesian optimization, including checkpoint support for model training. \\
\hline
Scalability&KGTK&\cite{ilievski2020kgtk}& A knowledge graph toolkit aimed at manipulating, deriving and enhancing KGs on a large scale. It provides the KGTK file format using the concept of hyper graphs for effective representation of information.\\
\cline{2-4}
& Marius &\cite{mohoney2021marius}& A framework built to optimize the data movement using partition caching and BETA data ordering for scalable training. \\
\cline{2-4}
& RDF2Vec Light &\cite{portisch2020rdf2vec}& An embedding approach built on RDF2VEC to selectively select a subset of triples to generate vectors using a walk generation algorithm.\\
\cline{2-4}
& Cleora &\cite{rychalska2021cleora}& An unsupervised learning approach based on optimizing the embeddings generated by the weighted ensemble of each node locality.\\
\hline
Dynamic Knowledge & NODE &\cite{ding2021temporal}& Exploits Neural ordinary differential equations and GNN to capture temporal and structure knowledge respectively.\\
\cline{2-4}
&CTDNE&\cite{nguyen2018dynamic}& Augmentation of random walk based models to capture the order of edges that moves ahead in time.\\
\cline{2-4}
& SDG &\cite{fu2021sdg}& Replace message passing mechanism in GNN by page rank mechanism to capture the dynamic nature in graphs. \\
\hline
Knowledge Quality & DSKRL &\cite{shao2021dskrl}& Proposed dissimilarity and support method for measuring the degree of similarity with its reliability using structural and
supporting information.\\
\cline{2-4}
& KGRefiner &\cite{saeedizade2021kgrefiner}& Augmenting current nodes by using their hierarchical information for a translation based embedding model to produce more informative graphs.\\
\hline
Knowledge Transfer& Pretrain-KGE &\cite{zhang2020pretrain}& A universal training framework with three steps namely, fine tuning, feature extraction and training phase using BERT\\

\hline
\end{tabular}
\end{center}
\end{table*}

The knowledge graphs address information in graph components and their associations. The prerequisites for KG emerge in view of the web improvements associated with the advancement of data on the Internet. Despite the various advantages of state-of-the-art KGs, they manage several issues. As shown in Table 3, many open search challenges may choose future assessment titles on knowledge development, refinement, reproducibility, etc. The given subsections have their share of challenges and issues.

\subsection{Interpretability and Robustness}
One of the primary functions of the KGE model is to embed the entities and relations of the KG into a low-dimensional vector space that is semantically meaningful. The embedding learning strategy works primarily by transforming the given vectors into auxiliary embeddings, which use gradient descent at a particular target objective loss. These systems, in any case, fill in as a black box, which is difficult to understand, unlike other approaches based on association rule mining and graph features, which can be deciphered dependent on the highlights they use. Preliminary efforts have been made to make the models more robust for link prediction tasks \cite{pezeshkpour2019investigating} \cite{bhardwajpoisoning} \cite{zhou2019adversarial}  \cite{xu2021robustness}. In this way, further work must go into interpretability and work on the reliance of the anticipated information and make the models more robust towards adversarial modifications.
 
\subsection{Few Shot Learning}
The relationships between the elements in the KG are far from complete, especially for unusual relationships, making it incredibly difficult to capture hidden examples of these relationships. Few shot learning is a technique proposed for learning in case the training data is low, which has previously been shown to have significant performance in image vision tasks \cite{fei2006one}. Few attempts have been made to incorporate few-shot learning strategies into the knowledge graph, aimed at finding hidden examples of a relation with which only certain triples are related \cite{ xiong2018one}  \cite{zhang2020few} \cite{du2019cognitive} \cite{mirtaheri2020one} \cite{hu2021can}. Although these are acceptable efforts, their low performance suggests that few-shot learning in graphs is an open research area.

\subsection{Hyper Parameter Search}
The resultant predictive exactness of KGE embeddings is highly dependent on their hyperparameters \cite{kadlec2017knowledge} \cite{ruffinelli2019you}. Hence, minor changes in these boundaries can negatively affect the predictive power of KGE models. The way towards tracking down the ideal boundaries of KGE models is customarily accomplished through a brute force parameter search which is time-consuming and inefficient. Thus, their preparation may require a rather novel framework that can optimize the training parameters to segment the optimal search space for each new dataset.

\subsection{Scalability}
Scalability is imperative in a large-scope knowledge graph. With a predetermined number of functions applied to over 1 million elements, there is a compromise between computational effectiveness and model expression. Some embedding strategies use rearrangement to reduce computing costs, for example, HolE\cite{nickel2016holographic}. ExpressGNN \cite{zhang2020efficient} tries to use NeuralLP \cite{yang2017differentiable} for efficient rule induction. Nonetheless, there is a need for a more thorough and tailored incorporation of Big Data development frameworks and modern factual models, and this remains an open research area. 

\subsection{Dynamic Knowledge}
It targets learning new rationale and interpreting new knowledge developing with time. Existing representation techniques are totally given to reasoning in the static KGs; however, they overlook the dynamic data contained in graphs. Of course, the triples contained in KGs, for example, (Steve Ballmer, CEO of, Microsoft) are not in every case valid over time. Plus, new information is created constantly, which might be infused into KGs progressively. In this manner, dynamic reasoning upon the huge KGs is requested to self-right KGs and mining new rationale techniques consistently.  A dynamic information graph, along with methods catching dynamics, may address the restriction of conventional information reasoning and representation by mining the temporal facts.
 
\subsection{Knowledge Quality}
KGE models form vector embeddings of elements according to their prior information. Therefore, the nature of this information affects the nature of the embeddings created. In fact, it is important to refine the information in KG to maintain its uniformity and accuracy. Refinement is an interaction between the inclusion of missing information and the identification of errors. KG refinement strategies \cite{paulheim2017knowledge} distinguish misinformation and allow congruent knowledge.

\subsection{Knowledge Transfer}
Neural reasoning techniques, for example, TransE and ConvE, construct the semantically meaningful embeddings of the facts. The generated embeddings can be transferred to the neuro-symbolic reasoning cycle to work on the limit of adaptation to fault tolerance. This reasoning cycle only aims to learn the features as boundaries present in the KG. In simple terms, it cannot be further transferred to some different KG. Lately, the graph neural networks (GNN) have successfully proven to capture the structure knowledge in graphs and transfer this knowledge to other graphs \cite{qiu2020gcc} \cite{hu2020gpt}. Roused by the accomplishment of the GNN pre-trained models, the KGE pre-trained model that can catch the adaptable semantics of the entities and relations across various datasets is an open research challenge.

\section{Applications}

Multi-relational graphs, also known as knowledge graphs, have substantially impacted several applications. Their property of delivering semantically organized data has brought important potential answers to some problems and provides excellent solutions. Many works devoted to leveraging KG have set their leanings on dialogue frameworks, recommendation frameworks, and information retrieval systems. Nevertheless, there are certain areas where KG's has wide applications in clinical, monetary, online security, news, and human resources.

\subsection{Information Retrieval Systems}
Recently, more and more web-based search frameworks use KG's fusing rich semantic entity information to help improve query results. It extends web indexing by using complex multi-relational information on real records, enriching queries and improving their ability to understand records. For the most part, search frameworks are well organized units to build with improvements over the vast range of knowledge graphs. Experts are investigating the potential of KG for information retrieval in several possible approaches that take advantage of KG's semantics in various segments, for example, query retrieval, document retrieval, and ranking models \cite{xiong2017word} \cite{liu2015latent}. Queries can be advanced and expanded by introducing new possibilities from related entities and their attributes, which then improve the function of query retrieval. The authors of \cite{dalton2014entity} used this idea by separating features from the elements themselves, and the relationships between entities to information bases, such as organized facts and text, used to advance the query. In document retrieval tasks, an approach to improve and address documents is to create a vector space model of the given elements \cite{raviv2016document}. As presented in \cite{raviv2016document}, a bag of entity vector space models is introduced in which documents and queries can be addressed using entity information. The authors of \cite{ensan2017document} proposed an entity linking framework to represent documents and questions as a set of semantic ideas, which can then be acted upon downstream tasks. The positioning vector model can also be improved by building various associations from queries to documents through related elements. 

\subsection{Recommender Systems}

It has been observed that the amount of online information like pictures, news and items is expanding, which brings confusion and issues for the users. Recommendation frameworks reduce the data overload faced by people emerging in this decade. These frameworks typically rely on collaborative filtering strategies, which dissect customers' prehistoric data and preferences. In any case, the primary impediment of exploiting this approach is that it experiences the sparsity of users' information and is computationally costly to prepare. 
The idea of side data can be used by recommendation systems by adopting knowledge graphs that deal with the issues mentioned earlier. KG can help improve accuracy, interpretability, and increase the variety of things recommended. 

KGE models are utilized to preprocess the KG by displaying the learned entity embeddings to a recommendation framework \cite{wang2018dkn} \cite{zhang2016collaborative} \cite{wang2019multi}. Hongwei Wang et al. \cite{wang2018dkn} proposed a content-based deep knowledge aware network (DKN) for news recommendation. The work \cite{zhang2016collaborative} developed a collaborative knowledge-based embedding (CKE) model by extracting semantic representation of items from past knowledge leveraging TransR. It considers the heterogeneity of both nodes and relationships, textual content, and visual content by utilizing stacked denoising and convolutional auto-encoders. A multi-task feature learning approach (MKR) is introduced in \cite{wang2019multi} for augmenting the recommendation leveraging knowledge graph embeddings. KGE based strategies have high adaptability in recommendation systems; however, the significant disadvantage is that they have no side data other than text. To provide additional information for recommendation, graph algorithms and path-based methods have also been employed to track significant associations between nodes in a knowledge graph \cite{zhao2017meta} \cite{wang2019explainable}. The above strategies are acceptable in making KG more general and intuitive to use. However, manual meta path feature extraction is generally troublesome and difficult to optimize practically. These techniques are also impossible where entities are relationships are not inside one realm like news suggestions.

\subsection{Questions Answering Systems}

Multi-relational graphs can be employ to upgrade indexed results to what is referred to as a question-and-answer (QA) system. For example, a QA framework, 'Watson', is built by IBM using YAGO \cite{suchanek2008yago} and DBpedia \cite{bizer2009dbpedia}. The QA systems are divided into types namely, semantic-based, information retrieval-based, embedding-based, and deep learning-based \cite{zou2020survey}. In a semantic-based QA framework, the semantics of the query can be communicated by turning standard language-based questions into logic structures. Then organized questions are prepared to elicit answers through the knowledge graph \cite{berant2013semantic} \cite{fader2014open}. Semantic parsing strategy shows excellent performance when managing complex queries. But, it relies on vast highlights or features being hand-made for semantic parsers, which restricts the application areas and versatility of their technology.

QA noting frameworks that are information retrieval based aim to automatically interpret given inquiries into organized questions to retrieve the arrangement of candidate answers from multi-relational graphs. Then, the attributes are extracted from the questions, and candidates can identify the correct answers and rank them. Information retrieval-based techniques rely more on natural language semantics and help manage basic queries. When compared with semantic and data retrieval strategy, the embedding-based model yields good results without hand-generated features or additional frameworks for grammatically labeling, dependencies, and syntactic parsing during preparation. Nonetheless, it overlooks word order data and cannot handle convoluted inquiries \cite{yao2014information}. 

Since the advent of deep learning models for natural language processing tasks, researchers have been attempting to take advantage of neural nets to perform QA tasks. It aims to reduce the over-reliance on hand-crafted features, which are time-consuming. A multi-channel CNN method for retrieving information is proposed \cite{dong2015question}. The concept of embedding and information retrieval techniques was adopted to reduce semantic parsing for query graph generation \cite{yih2015semantic}. Bidirectional LSTM is applied by Zhang et al. \cite{zhang2016question} to get familiar with the representations of queries leveraging embedding-based methods.

Regular QA framework can be seen as a single round response framework by formulating correct answers in the form of feedback. However, dialogue frameworks are in advance because of their potential to quickly create multi-round responses through semantic enhancements and KG walks. An encoder–decoder structure-based graph attention component has been proposed by Liu et al. \cite{liu2019knowledge} to encode the data to enhance the semantic representation. The literature proposed \cite{moon2019opendialkg} to learn logical progressions in turn via representative knowledge graphs navigating to the anticipated response with the attention graph path-based decoder. Semantic parsing through formal logical representation is another heading for dialogue frameworks \cite{guo2018dialog}; in any case, they are difficult to parse and interpret.

\subsection{Other Applications}

\subsubsection{Health Informatics}
Clinical data is evolving rapidly, and natural language information is quite common and occupies an important place in the health informatics framework.  Endeavors have been made to use the accessible data into knowledge graphs to furnish frameworks with extricating and compiling clinical information accurately and speedily. A strategy for creating a huge-scope biomedical KG was proposed by Ernst et al. \cite{ernst2014knowlife}. The health information was effectively coordinated into a heterogeneous literary information graph in \cite{shi2017semantic}. A method has been proposed by Rotmensch et al. \cite{rotmensch2017learning} to effectively exploit the knowledge in electronic medical records to map diseases with symptoms automatically. The previously mentioned approaches help build large-scale knowledge graphs taking into account the standard clinical terminology. But, especially for dialects like Chinese, the standards are not necessarily met. This brings about the relatively low accuracy of clinical KG in such dialects.

\subsubsection{Drug Discovery}
Drug development is a difficult and costly cycle, from gene distinguishing evidence to quality checks and identifying a compound for experimentation on subjects. Inherent progress of a gene or drug requires many years and can result in loss of time and resources if not identified effectively. Drug developers identify genes and drugs by reading the most recent literature before continuing with the experiment. In any case, it is profoundly reliant upon the experience of the specialists. Knowledge graph embeddings can be used to deal with these issues \cite{mohamed2020discovering} \cite{mohamed2019drug}. A knowledge graph can be created based on a genetic approach by combining different genes and their associations for a particular disease. The KGE's can then be employed to learn complex interactions from graphs and make link predictions to predict associations present in the dataset. It will follow some priority-based ranking protocols and list the priority genes in hierarchical order of calibration. The drug designer may then discover the confirmation behind the expected results and continue working accordingly.

\subsubsection{Covid-19}
The current worldwide emergency brought about by COVID-19 nearly stopped normal life in many places of the world. As of September 1, over 220 million individuals were infected, and the quantity of COVID-19 patients is drastically expanding. The shortage of medical supplies is at its peak and has likewise turned into a significant challenge \cite{okereke2021impact}. Knowledge Graphs (KG) have been demonstrated to effectively look through the overwhelming volume of COVID-19 literature and gain actionable understanding, which would either be very monotonous or difficult to accomplish without leveraging AI. The applications are comprehensively sorted into primary parts, i.e., in Drug Repurposing and Knowledge Graph Construction \cite{chatterjee2021knowledge}. In Drug repurposing, the task is to locate potential medications to repurpose for COVID-19 utilizing literature determined information and KGC methods \cite{al2021knowledge}. The latter task involves building a knowledge graph to facilitate literature search \cite{kim2021deep}.

\subsubsection{Human Rescource}
Technology is advancing at an astonishingly high speed, and job seekers need to acquire new capabilities to be relevant in the marketplace. But due to automation, many jobs are becoming obsolete, and organizations are forced to lay off people. The KGE's can be used to propose new technologies or tasks to professionals and suggest comparable roles and skills within the organization \cite{zhang2019job2vec}. It can also estimate the relationship between entities in monetary business areas and deduce recognized patterns in knowledge.

\subsubsection{Knowledge Protection}
Advances in information technology are at their peak, resulting in the need to prevent cyber attacks from securing the system fully. Increasingly more examination work is done leveraging KG's associated with cyber security to identify and anticipate dynamic attacks and protect individuals' data. A five-level model introduced by Jian et al. \cite{jia2018practical} is built on a network security knowledge graph that aims to obtain updated knowledge using path ranking algorithms. The study \cite{qi2018association} shows patterns associated with digital attacks and breaks down links between attacks, incidents, and precautions by elevating the nature of the incident. 
The given approaches are more focused on the development of data security KG. Nevertheless, how to isolate digital security incidents using KG's indistinguishable information thought capability and rapidly update KG with new improvements by researchers requires further exploration in the future.

\section{Conclusion}

Knowledge Graph Completion (KGC) is an intriguing issue in KG development and related applications, which expects to construct KG by foreseeing the missing relationships and entities. The thought behind introducing a paper like this was to conduct an examination that assembles all cutting-edge approaches on the knowledge graphs for link prediction.  To the best of our knowledge, this paper is one of only a handful of works that systemically give an outline on knowledge graph completion from various methodologies such as conventional statistical relational learning, probabilistic graphical models, rule mining, and graph representation learning techniques. 

Knowledge graph embeddings (KGE), as the innovation of feeding triples into a low-dimensional consistent vector space, has gained astounding headway in offering exact, viable, and underlying portrayal of data in numerous fields. We explored primary advances of KGE, sorted the current scoring capacities into three kinds, namely, translation, factorization, and deeper scoring functions, and afterward outlined the benefits and weaknesses of embedding models in every class. Techniques based on reinforcement learning like DeepPath, Minerva, M-Walk, DAPath, and many more are also reviewed and discussed to estimate complex queries in the link prediction task.

When seeing examination papers and conversing with specialists in link prediction, we discovered a slight disconnect between what analysts accept best in class and what really is cutting edge. Along these lines, a comparative analysis is given to help fellow researchers to comprehend the less explored areas like calibration, reciprocal relations, and different training strategies. Our work is efficient, simple to study, and contains detailed figures and tables. A bird's eye view of the current state-of-the-art models on KGC for temporal, uncertain, and multimodal knowledge graphs is also discussed. Next, we compared recently published software packages for model training and open research challenges, also discussed to guide future research. 

We firmly accept that this literature will help researchers influence our findings and act as stepping stones to push the cutting edge. Even though, there are a few impediments in this work because of space and time constraints. This study centered around the KGE for link prediction; we will investigate more research areas of Knowledge graph completion in later versions, like triple classification, entity classification, etc. Additionally, we emphasize static KG; we will explore new model designs, like dynamic, heterogeneous, and bipartite graphs.

\bibliographystyle{model1-num-names}
\bibliography{sample.bib}







\end{document}